\documentclass{emulateapj}

\usepackage{graphics}
\usepackage{epsfig}
\usepackage{natbib}
\shorttitle{CO/H$_{2}$ in the RW Aur A Disk}
\shortauthors{France et al.}
\begin{document}

\title{CO/H$_{2}$ Abundance Ratio $\approx$~10$^{-4}$ in a Protoplanetary Disk\altaffilmark{*}}


\author{
Kevin France\altaffilmark{1}, Gregory J. Herczeg\altaffilmark{2}, Matthew McJunkin\altaffilmark{1}, Steven V. Penton\altaffilmark{3}	
}

\altaffiltext{*}{Based on observations made with the NASA/ESA $Hubble$~$Space$~$Telescope$, obtained from the data archive at the Space Telescope Science Institute. STScI is operated by the Association of Universities for Research in Astronomy, Inc. under NASA contract NAS 5-26555.}

\altaffiltext{1}{Center for Astrophysics and Space Astronomy, University of Colorado, 389 UCB, Boulder, CO 80309; kevin.france@colorado.edu}
\altaffiltext{2}{ Kavli Institute for Astronomy and Astrophysics, Peking University, Beijing 100871, China}
\altaffiltext{3}{Space Telescope Science Institute, 3700 San Martin Drive, Baltimore MD, 21218, USA}


\begin{abstract}
The relative abundances of atomic and molecular species in planet-forming disks around young stars provide important constraints on photochemical disk models and provide a baseline for calculating disk masses from measurements of trace species.  A knowledge of absolute abundances, those relative to molecular hydrogen (H$_{2}$), are challenging because of the weak rovibrational transition ladder of H$_{2}$ and the inability to spatially resolve different emission components within the circumstellar environment.  To address both of these issues, we present new contemporaneous measurements of CO and H$_{2}$ absorption through the ``warm molecular layer'' of the protoplanetary disk around the Classical T Tauri Star RW Aurigae A.  We use a newly commissioned observing mode of the {\it Hubble Space Telescope}-Cosmic Origins Spectrograph to detect warm H$_{2}$ absorption in this region for the first time.  An analysis of the emission and absorption spectrum of RW Aur shows components from the accretion region near the stellar photosphere, the molecular disk, and several outflow components.  The warm H$_{2}$ and CO absorption lines are consistent with a disk origin.  We model the 1092~--~1117~\AA\ spectrum of RW Aur to derive log$_{10}$ N(H$_{2}$)~=~19.90$^{+0.33}_{-0.22}$ cm$^{-2}$ at T$_{rot}$(H$_{2}$) ~=~440~$\pm$~39 K.  The CO $A$~--~$X$ bands observed from 1410~--~1520~\AA\ are best fit by log$_{10}$ N(CO)~=~16.1~$^{+0.3}_{-0.5}$ cm$^{-2}$ at  T$_{rot}$(CO) ~=~200$^{+650}_{-125}$ K.   Combining direct measurements of the \ion{H}{1}, H$_{2}$, and CO column densities, we find a molecular fraction in the warm disk surface of $f_{H2}$~$\geq$~0.47 and derive a molecular abundance ratio of CO/H$_{2}$~=~1.6$^{+4.7}_{-1.3}$~$\times$~10$^{-4}$, both consistent with canonical interstellar dense cloud values.  
\end{abstract}


\clearpage

\section{Introduction}

The gas and dust in protostellar disks provide the raw materials for planet building.   The formation of giant planet cores through the coagulation of dust grains~\citep{hayashi85} is thought to be complete prior to the 2~--~4 Myr dust disk clearing timescale~\citep{hernandez07,ingleby11b,fang13}.  The majority of giant planet formation is thought to take place inside of ~$\sim$~10 AU~\citep{mordasini09}, and these protoplanets accrete their outer layers and atmospheres from the protoplanetary gas disk prior to its dissipation~\citep{ida04}.  The final mass and composition of protoplanets are therefore closely related to the abundances, spatial distributions, and lifetimes of the gas in the circumstellar environment.  The gas disk also regulates planetary migration~\citep{ward97,armitage02,trilling02}; the migration timescale is sensitive to the specifics of the disk surface density distribution and lifetime~\citep{armitage07}.  Gas disk dissipation timescales inferred from accretion indicators are found to be similar to the dust-clearing timescale ($\approx$~2~--~5~Myr; Fedele et al. 2010; Jayawardhana et al. 2006; Sicilia-Aguilar et al. 2005).   However, direct gas disk observations indicate that inner molecular disks can persist to ages $\gtrsim$~10 Myr in some Classical T Tauri Stars (CTTSs) and transitional systems (see, e.g., the review presented in Najita et al. 2007; Salyk et al. 2009; Ingleby et al. 2011a; France et al. 2012b), although these results are based on a relatively small number of protoplanetary systems.~\nocite{salyk09,ingleby11a,france12b,fedele10,rayjay06,aguilar05,najita07}

The composition of a planetary system is also impacted by the initial abundances in the protoplanetary environment, in particular the C/O ratio~\citep{bond10,oberg11}.  Observations and models of the atomic and molecular composition of young circumstellar disks are useful tools for creating an inventory of the materials available for planet formation (e.g., Aikawa et al. 1997; Thi et al. 2004), as well as providing constraints on photochemical models of the protoplanetary environment~\citep{dullemond07,visser11}.~\nocite{aikawa97,thi04}
An unexpected observational discovery from the {\it Hubble Space Telescope}-Cosmic Origins Spectrograph (COS) was the detection and characterization of carbon monoxide (CO) in the far-ultraviolet (far-UV; 1150~$\leq$~$\lambda$~$\leq$~1750~\AA) spectra of low-mass protoplanetary systems. \citet{france11b} presented the first detections of far-UV emission and absorption lines of CO in these environments.  It was shown that these CO lines provide unique diagnostics of the disk structure and that the strength of these features is surprising in light of the expected abundance of CO in the disk.  Models of the CO and H$_{2}$ emission indicated that the observed CO/H$_{2}$ ratio ($\equiv$~$N$(CO)/$N$(H$_{2}$)) was in the range  0.1~$\leq$~CO/H$_{2}$~$\leq$~1~\citep{france11b,schindhelm12a}.

Initial CO and H$_{2}$ absorption line measurements in CTTSs indicated similarly high CO/H$_{2}$ ratios.  ~\citet{france12a} presented an analysis of the molecules on a sightline through the AA Tauri circumstellar disk, observing CO against the far-UV continuum and H$_{2}$ absorption against the broad Ly$\alpha$ line that originates in the protostellar atmosphere (see also Yang et al. 2011).\nocite{yang11}  In agreement with the emission line work, they find CO/H$_{2}$ $\approx$~0.4, providing an independent measure of large CO abundance ratios in protoplanetary environments.

The large abundances of CO are surprising because protoplanetary disks form in dense clouds, where the CO/H$_{2}$ ratio is usually assumed to be $\sim$~10$^{-4}$~\citep{lacy94}.  Furthermore, recent work on the gas composition at larger disk radii suggests a depleted CO/H$_{2}$ ratio~\citep{bergin13,favre13}.    
These results raise questions about the mass budget and chemical composition of the gas phase at planet-forming radii ($a$~$<$~10 AU).   Do the UV data imply that the local CO/H$_{2}$ abundance ratio of order unity, or is this result biased by spatial stratification of the emitting/absorbing molecular populations?  Because we expect H$_{2}$ to be abundant in almost all regions of the protoplanetary gas disk, is there a particular spatial or temperature structure that makes the majority of the warm H$_{2}$ hard to detect?  From an observational perspective, the question is: where is the H$_{2}$ that should be associated with the large reservoir of CO observed in the UV spectra?  

\subsection{CO/H$_{2}$ at Planet-forming Radii, Where is the Warm H$_{2}$?} 
Despite considerable observational effort dedicated to the 
characterization of warm H$_{2}$ ($T_{rot}$(H$_{2}$)~$\sim$~500 K) in CTTS environments, the rovibrational emission lines have proven challenging to characterize in sources without strong outflows (e.g., Beck et al. 2008).   The homonuclear nature of H$_{2}$ means that rovibrational transitions are dipole forbidden, with weak quadrupole transitions that have large energy spacing.\nocite{pascucci06,carmona08,bary08,zaidi10,lahuis07} This makes direct detection of H$_{2}$ challenging at near- and mid-IR wavelengths (Pascucci et al. 2006; Lahuis et al. 2007; Bitner et al. 2008; but see also Bary et al. 2008), and dedicated searches have usually returned upper limits (e.g., Carmona et al. 2008; Martin-Za{\"i}di et al. 2010) or tentative detections around more massive Herbig Ae star disks~\citep{richter02}.

\begin{figure}
\begin{center}
\epsfig{figure=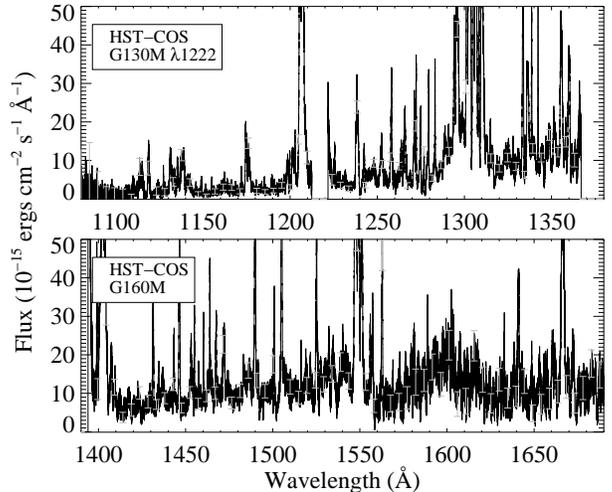,width=2.65in,angle=90}
\vspace{-0.05in}
\caption{
\label{cosovly} Overview of the 29 August 2013 $HST$-COS observations of RW Aur A.  The data are displayed as the black histogram with representative error bars shown in gray.  
 }
\end{center}
\end{figure}

The observational results described above are roughly consistent with disk models that find a significant CO population in the disk surface near N$_{H}$~$\sim$~10$^{21}$ cm$^{-2}$ where the hydrogen molecular fraction is low (e.g., Glassgold et al. 2004); in this case the regions of high local CO/H$_{2}$ ratio may be limited to narrow layers in the upper disk atmosphere where far-UV H$_{2}$ fluorescence originates, and not representative of the warm molecular disk as a whole.   A key difference between the CO and H$_{2}$ populations observed in UV spectra is the derived rotational temperature: CO emission/absorption originates in a warm ($T_{rot}$(CO)~$\sim$~300~--~700 K, $N$(CO)~$\sim$~10$^{17-19}$ cm$^{-2}$) molecular gas while the H$_{2}$ emission/absorption comes from a hotter ($T_{rot}$(H$_{2}$)~$\sim$~2000~--~3000 K, $N$(H$_{2}$)~$\sim$~10$^{18-19}$ cm$^{-2}$) molecular phase~\citep{herczeg04,schindhelm12a}.  One interpretation is that the previous UV observations probe CO at large semi-major axes ($a$~$\gtrsim$~2AU), and that this gas is spatially distinct from the hot H$_{2}$-emitting gas orbiting at terrestrial planet radii (0.1~$<$~$a$~$\lesssim$~2 AU).  This interpretation is supported by an analysis of the emission line-widths.  Assuming that Keplerian rotation dominates the observed velocity broadening, the narrower spectral lines of the warm CO suggest an origin at larger disk radii than the hot H$_{2}$~\citep{schindhelm12a,france12b}.   Alternatively, if the line-broadening is not dominated by rotation and non-thermal processes control the molecular level populations, then the rotational temperature may not reflect the local kinetic temperature of the molecular gas~\citep{france12a}.   In this case, the observed CO and H$_{2}$ may be approximately co-spatial with a very large CO abundance. 

The combination of CO and H$_{2}$ fluorescence observations and emission line modeling have raised intriguing questions regarding the composition and spatial distribution of the molecular material at planet-forming radii, however, absorption spectroscopy provides the most direct, model-independent means of measuring the column densities on the line-of-sight through these disks.  
An analysis of 34 T Tauri stars found a roughly 25\% detection rate of warm CO absorption similar to AA Tau in gas-rich disks~\citep{mcjunkin13}.  The McJunkin study identified six moderate-inclination 
disks with $T_{rot}$(CO)~$\sim$~500 K CO absorption.  These targets spanned a range of stellar masses ($\sim$~0.4~--~2.3~M$_{\odot}$), ages ($\sim$~0.6~--~6~Myr), and mass accretion rates ((0.1~--~3)~$\times$~10$^{-8}$ M$_{\odot}$ yr$^{-1}$).    While H$_{2}$ absorption spectroscopy has been carried out for the intrinsically hotter and brighter Herbig stars~\citep{roberge01,zaidi08} by the {\it Far-Ultraviolet Spectroscopic Explorer} ($FUSE$), lower-luminosity CTTSs do not produce sufficient flux to be used as a background source for disk absorption studies with $FUSE$.   
%

The McJunkin et al. sample took advantage of the increased sensitivity of the $HST$-COS to provide a target list of protoplanetary disks with known molecular absorbers, but relatively low reddening.  
 Unfortunately, the ``traditional'' $HST$ far-UV bandpass (1150~--~1750~\AA) does not provide spectral coverage of H$_{2}$ gas with kinetic temperatures of a few hundred degrees.  Thermal excitation at 300~--~700 K will produce an appreciable population of H$_{2}$ in the $v$~=~0, $J^{''}$~=~0~--~5 levels, whose transitions reside at $\lambda$~$<$~1126~\AA\ (the longest wavelength transition for H$_{2}$ in $J^{''}$~=~5 is the Lyman (0~--~0) P(5) line at 1125.54~\AA).   Spectral coverage in the 1090~$\lesssim$~$\lambda$~$\lesssim$~1130~\AA\ wavelength region is required to measure both the column density and kinetic temperature of the warm H$_{2}$ disk.  

During on-orbit verification following $HST$ Servicing Mission-4, it was demonstrated that $HST$~+~COS maintains spectroscopic sensitivity down to the Lyman edge at 912~\AA~\citep{mccandliss10}.   We take advantage of this short-wavelength response, in combination with a new medium resolution mode of COS developed in part for this work, to directly measure the CO/H$_{2}$ column density ratio in the warm molecular phase of a CTTS disk, RW Aur A (Section 2), for the first time.  In Section 3, we describe the new G130M $\lambda$1222 mode used for these observations of the disk around RW Aur A and the data analysis used to extract H$_{2}$ and CO absorption line profiles from the data.  In Section 4, we describe the CO and H$_{2}$ modeling analyses used to derive column densities and excitation temperatures in RW Aur A, and demonstrate that this line-of-sight absorption originates in the circumstellar environment.  Section 5 presents an analysis of the velocity fields present in RW Aur A, showing that the molecular disk component can be readily separated from the molecular and atomic outflows.  We conclude Section 5 with a discussion of the CO/H$_{2}$ ratio and present a brief summary in Section 6.

\section{RW Aurigae}

The RW Aur system is composed of two pre-main-sequence K stars, separated by approximately 1.5\arcsec~\citep{duchene99}, at a distance from Earth of $d$~$\approx$~140 pc~\citep{elias78,kenyon95,loinard07}.  The primary component (RW Aur A) is likely a K0~--~K4 star ($M_{\star}$~$\approx$~1.1~--~1.4~M$_{\odot}$; see Woitas et al. 2001 and references therein), roughly 30~--~50\% more massive than the K5~--~K7 B component~\citep{herczeg14}.~\nocite{woitas01}   RW Aur A displays a near-infrared (near-IR) excess indicative of a warm inner dust disk~\citep{hartigan95} and a total disk mass (assuming a gas-to-dust ratio of 100)  $\sim$~4~$\times$~10$^{-3}$~M$_{\odot}$~\citep{andrews05}. 

Disk inclination estimates range from 45\arcdeg~--~90\arcdeg, with sub-mm maps favoring lower inclinations~\citep{cabrit06} and near-IR interferometry suggesting higher values~\citep{eisner07}.  \citet{cabrit06} describe high-resolution single-dish interferometric measurements of the millimeter dust continuum and CO rotational lines.  Modeling these data as a Keplerian disk rotating about the jet-axis of the system (see below), they find best-fit outer disk inclinations ranging from 45~--~60\arcdeg, depending on assumptions about jet structure and velocity.  \citet{eisner07} describe multiple epochs of near-IR spectroscopy and interferometry to constrain the size, luminosity, variability and inclination of the $inner$ disk ($r$~$<$~2 AU) around RW Aur A.  Their observations and subsequent uniform disk modeling find $i$~=~77\arcdeg$^{+13}_{-15}$.  

CO fundamental emission lines observed at NIRSPEC~\citep{najita03} suggest a double-peaked emitting structure, with a best-fit disk inclination of $i$~=~60\arcdeg.  However, these observations and higher-resolution spectra from CRIRES~\citep{brown13} show that the broad CO lines are severely blended and may include a contribution from a molecular wind, complicating the estimation of Keplerian disk parameters from these data.  The 4.7~$\mu$m CO data only show CO emission, with no central reversal as is observed in some CTTS and Herbig spectra~\citep{brown13}.   McJunkin et al. (2013) combined simple disk structure models with the first epoch of $HST$-COS CO absorption line data (described above) to determine that the $A$~--~$X$ absorption lines originate above the A$_{V}$~=~1 dust surface, finding that for dimensionless disk height $z$/$r$~=~0.6, the RW Aur disk must have inclination~$>$~60\arcdeg\ to account for the observed absorption. In the subsequent sections, we will argue that the majority of the absorbing molecular gas that we observe in the RW Aur A system comes from a warm inner region.  Evaluating the various inclination estimates, we conclude that the inner disk inclination of RW Aur A is $i$~$\geq$~60\arcdeg.

Using a combination of ground-based optical spectra, $HST$ near-UV spectra, and accretion shock models, \citet{ingleby13} find a mass accretion rate of 2.0~$\times$~10$^{-8}$ M$_{\odot}$ yr$^{-1}$ for RW Aur A.   The more massive A component appears to have a higher mass accretion rate~\citep{hartigan95,white01} and therefore should dominate the far-UV accretion luminosity that creates the background flux for the disk absorption line spectroscopy studied here.  Furthermore, the $HST$-COS instrumental response falls off sharply at $\theta$~$>$~0.5\arcsec\ from the center of the primary science aperture, and we conclude that RW Aur B does not contribute significantly to the observed far-UV flux.  The optical extinction towards RW Aur A is not well constrained due to heavy veiling, but the optical extinction towards RW Aur B (A$_{V}$~=~0.10; Herczeg \& Hillenbrand 2014) is very similar to the value found from interstellar Ly$\alpha$ absorption (log$_{10}$ N(HI) = 20.25$^{+0.05}_{-0.21}$ cm$^{-2}$) towards RW Aur A, assuming standard ISM gas-to-dust conversions (A$_{V}$~=~0.11; McJunkin et al. 2014).   

RW Aur A has a well-studied bipolar outflow, with the redshifted component ($v_{red}$~$\sim$~+100 km s$^{-1}$) having higher density and surface brightness than the blueshifted component~\citep{hirth94,melnikov09}.   The far-UV spectrum of RW Aur A has been studied in detail in two recent papers, one focusing on hot gas emission originating near the stellar surface~\citep{ardila13} and one studying  H$_{2}$ emission from the disk and outflow~\citep{france12b}.  RW Aur showed the
highest velocity molecular outflow emission in the \citet{france12b} survey, with both redshifted and blueshifted emission observed in the strong H$_{2}$ fluorescence lines excited by Ly$\alpha$ photons produced in the outflow/jet, reminiscent of the spatially-resolved T Tau molecular outflow~\citep{walter03}.  The UV H$_{2}$ emission peaks 80~--~110 km s$^{-1}$ to the red of the stellar velocity, suggesting that the molecular emission arises in material that is approximately cospatial with the forbidden atomic line (e.g., [\ion{S}{2}] $\lambda$6731) emission~\citep{woitas02,melnikov09,hartigan09}.   The near-IR H$_{2}$ outflow from RW Aur is centered near~+44 km s$^{-1}$~\citep{beck08}, significantly bluer than the peak of the far-UV H$_{2}$ velocity profile.  We will return to the velocity structure of the far-UV spectrum in Section 5.1.  

\begin{figure}
\begin{center}
\epsfig{figure=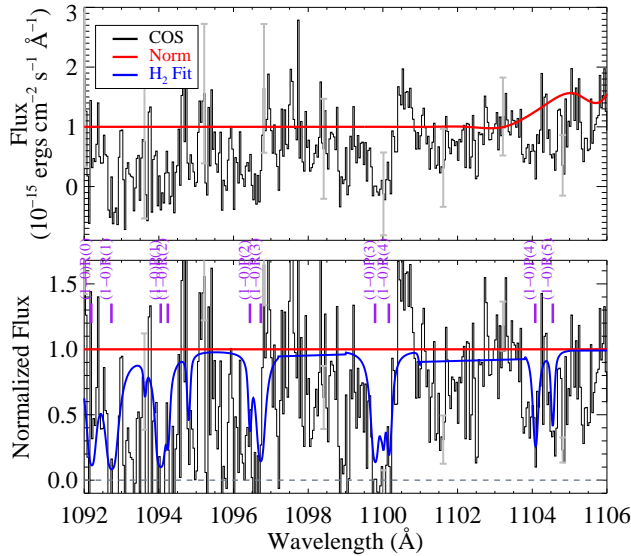,width=2.65in,angle=90}
\vspace{+0.1in}
\caption{
\label{cosovly} $HST$-COS spectra of the 1092~--~1106~\AA\ spectral region, with the continuum function shown as the solid red line (top panel).  The bottom panel shows the normalized flux and the best-fit H$_{2}$ absorption model as the blue solid line (\S 4.1).   The strongest H$_{2}$ absorption lines with $J^{''}$~$\leq$~5 are labeled in purple.  The data and the normalized spectra are binned by 4 pixels and smoothed with a 3 pixel boxcar average for display.  The normalized spectra are smoothed with a 3 pixel boxcar average prior to H$_{2}$ fitting.  
 }
\end{center}
\end{figure}

\begin{figure}
\begin{center}
\epsfig{figure=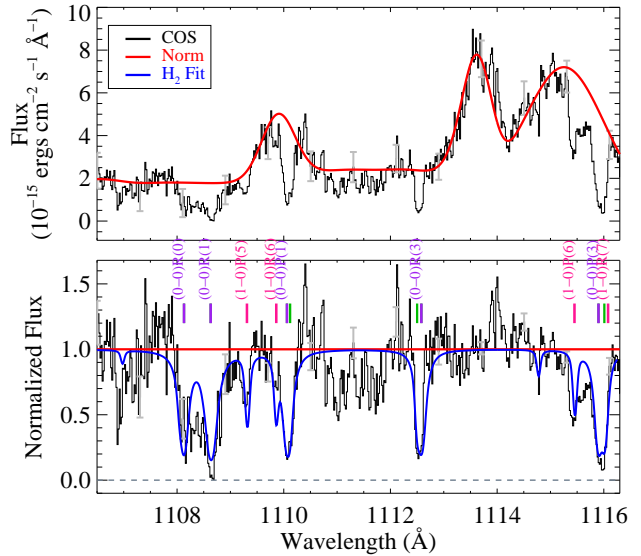,width=2.65in,angle=90}
\vspace{+0.1in}
\caption{
\label{cosovly} $HST$-COS spectra of the 1106~--~1116.5~\AA\ spectral region, with the continuum function shown as the solid red line (top panel).  The large emission features in the red spectrum are H$_{2}$ fluorescent emission lines, approximated by Gaussians for the purpose of constructing the normalization curve.  
The bottom panel shows the normalized flux and the best-fit H$_{2}$ absorption model as the blue solid line (\S 4.1).  The data and the normalized spectra are binned by 2 pixels and smoothed with a 3 pixel boxcar average for display.  The normalized spectra are smoothed with a 3 pixel boxcar average prior to H$_{2}$ fitting.  The best-fit H$_{2}$ absorption model is parameterized with $N(H_{2})$~=~19.90$^{+0.33}_{-0.22}$ cm$^{-2}$, $b_{H2}$~=~4~$\pm$~1 km s$^{-1}$, a radial velocity $v_{H2abs}$~=~5~$\pm$~5 km s$^{-1}$ and a covering fraction $f_{cov}^{H2}$~=~0.974~$\pm$~0.027.  Transitions to $v^{'}$~=~0 are labeled in purple and transitions to $v^{'}$~=~1 are labeled in pink.  For display clarity, we have omitted labels from close blends, noting them with a green tick.  The unlabeled lines are: (0~--~0) R(2) 1110.12~\AA, (0~--~0) P(2) 1112.50~\AA, and (0~--~0) R(4) 1116.01~\AA.  
 }
\end{center}
\end{figure}

\begin{figure*}
\begin{center}
\epsfig{figure=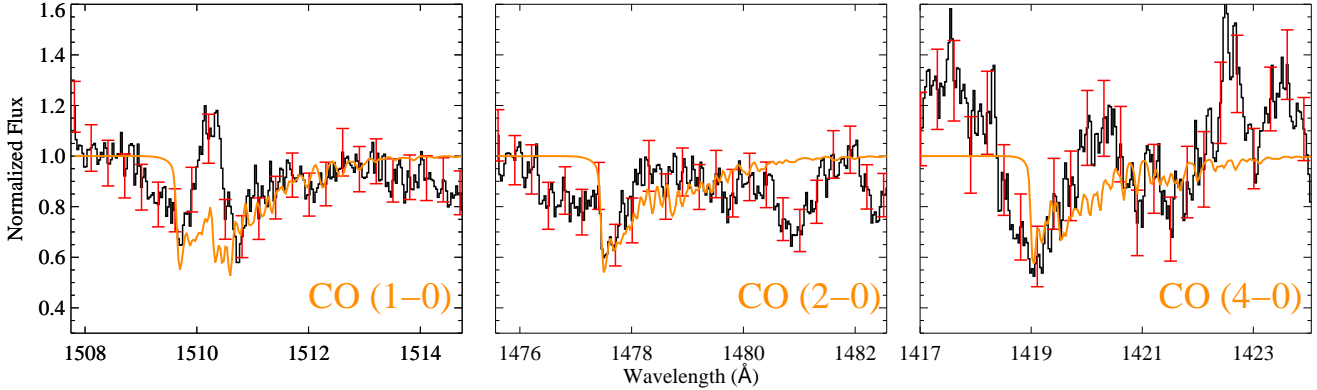,width=2.0in,angle=90}
\vspace{+0.1in}
\caption{
\label{cosovly} CO ($A$~--~$X$) absorption spectra for $v^{'}$~=~1, 2, and 4.  The data are normalized to the local continuum level around each band, and the best-fit model spectrum is shown in orange.  Contamination by H$_{2}$ emission lines compromise continuum and absorption band determination and drive the uncertainty on the best-fit CO absorption model, parameterized with $N$(CO)~=~16.1$^{+0.3}_{-0.5}$ cm$^{-2}$, $T_{rot}$(CO)~=~200$^{+650}_{-125}$, $b_{CO}$~=~0.5~$\pm$~0.1 km s$^{-1}$, and $v_{COabs}$~$\approx$~5~$\pm$~5 km s$^{-1}$.
 }
\end{center}
\end{figure*}

\section{Observations and Data Reduction: Direct Measurement of Warm Circumstellar H$_{2}$ at $\lambda$~$<$~1120~\AA\ with $HST$-COS}

The G130M $\lambda$1222 mode on COS, introduced for $HST$ Cycle 20 observations, has made possible high-sensitivity, moderate-resolution ($R$~$\geq$~10$^{4}$; $\Delta$$\lambda$~$\leq$~0.1~\AA) spectroscopy in the 1064~--~1130~\AA\ bandpass from $HST$ for the first time,  delivering $\approx$~15$\times$ the effective area of $FUSE$ at 1110~\AA.   
The {\tt CENWAVE} $\lambda$1222 mode is chosen by moving the angle and focus position of the G130M grating to a setting that is considerably beyond the original G130M default locations~\citep{penton12,penton13}.  The COS ``optics select mechanism 1'' is rotated $\approx$~0.8\arcdeg\ and the grating focus mechanism is moved by $\approx$~0.78 mm past the default ({\tt CENWAVE} $\lambda$1291) range.  
This setting allows COS observations over the spectral range 1064 -- 1214~\AA\ on the FUV ``B' segment, and 1223 - 1368~\AA\  on the FUV ``A'' segment. This configuration intentionally places geocoronal Ly$\alpha$\ on the small gap between the detector segments to minimize microchannel plate gain sag. The instrument focus in the $\lambda$1222 setting is chosen to optimize resolution on the FUV B-segment.

The observations presented here are the first science observations collected with the G130M $\lambda$1222 mode, obtained as part of the {\it Warm H$_{2}$ In Protoplanetary Systems} (WH$_{2}$IPS) Cycle 20 $HST$ Guest Observing program (PID 12876). 
The WH$_{2}$IPS observing strategy is to combine long exposure times (several orbits) at the short wavelength end of the COS bandpass to measure lines from the H$_{2}$ (1~--~0) $\lambda_{o}$~$\sim$~1092~\AA\ and (0~--~0) $\lambda_{o}$~$\sim$~1108~\AA\  Lyman band systems with shorter observations with the G160M grating (1400~--~1750~\AA) to contemporaneously observe the CO $A$~--~$X$ electronic absorptions into the 0~$\leq$~$v^{'}$~$\leq$~4 vibrational bands.  The CO absorption spectrum along the RW Aur A sightline was analyzed by~\citet{mcjunkin13}, however the system geometry and accretion rates of CTTS systems are time-variable, so we re-observed the CO spectra to be directly comparable with our new H$_{2}$ absorption line measurements.  

We observed RW Aur A (05$^{h}$ 07$^{m}$ 49.58$^{s}$ +30\arcdeg\ 24\arcmin\ 04.9\arcsec) with $HST$-COS on 29 August 2013.  We employed the G130M $\lambda$1222 mode for a total of 10468s (4 orbits) in all four focal-plane split positions~\citep{green12}.  We observed with the longer-wavelength G160M mode in four central wavelength settings, each with a different focal-plane position for a total of 5613s (2 orbits).  The use of multiple focal-plane positions allows for continuous wavelength coverage while minimizing the impact of fixed-pattern noise.  We coadded the spectra using a modified version of the IDL-based COS far-UV data reduction routines first described by~\citet{danforth10}.   The data, displayed in Figure 1, have a spectral resolving power of~$R$~$\approx$~16000 ($\Delta$$v$~=~19 km s$^{-1}$) at 1110~\AA\ and $R$~$\approx$~18000 ($\Delta$$v$~=~17 km s$^{-1}$) at 1600~\AA.  The absolute velocity accuracy of
these modes is $\approx$~15 km s$^{-1}$.

The primary scientific focus of this work is the quantitative treatment of the molecular lines observed in absorption against the far-UV continuum from RW Aur A.  The far-UV continuum in CTTSs is a superposition of a molecular quasi-continuum (1400~$\lesssim$~$\lambda$~$\lesssim$~1660~\AA) and a component that can be well-represented by a linear or quadratic function~\citep{france11a}.  The linear continuum is well-correlated with accretion indicators such as \ion{C}{4} line flux, however appears distinct from the well-studied near-UV Balmer continuum~\citep{france14}.  In any case, the far-UV continuum is smoothly varying on spectral scales of $\sim$~10~\AA, and we assume the local continua have a linear shape for the purposes of continuum normalization (Figure 2).  

 Because the COS observations are spatially unresolved, we are observing the entire star-disk system simultaneously.  As such there are many fluorescent H$_{2}$ emission lines that contaminate the fitting regions.  In the region of the H$_{2}$ $B$$^{1}\Sigma^{+}_{u}$~--~$X$$^{1}\Sigma^{+}_{g}$ (0~--~0) band between 1108~--~1116~\AA, there are several blended H$_{2}$ emission lines upon which we observe the H$_{2}$ absorption (Figure 3).  We normalize these spectra by assuming that the underlying spectrum can be parameterized with a combination of linear continuum and Gaussian emission lines, as shown in Figure 3.  For the case of the CO absorption spectra in the G160M band (Figure 4), the emission lines are typically narrow and we simply exclude these regions when computing the best-fit CO parameters.

\section{Column Density and Temperature Analysis}

\subsection{H$_{2}$ Fitting}

In order to assess the CO/H$_{2}$ abundance ratio in the RW Aur disk, as well as determine the co-spatiality of the absorbing gas,  the column densities, rotational temperatures, and velocity structures must be derived from the spectra.  The H$_{2}$ absorption lines were fitted with an IDL-based routine that combines the $H_{2}ools$ optical depth templates (McCandliss 2003) and the MPFIT least-squares minimization routines~\citep{markwardt09}.\nocite{mccandliss03}  The column densities of each rotational level $J^{''}$ in the ground electrovibrational level, $N$(H$_{2}$[$v^{''}$~=~0,$J^{''}$]) (in units of cm$^{-2}$), are allowed to vary simultaneously.  The first 9 rotational levels are considered ($J^{''}$~=~0~--~8); lines originating in $J^{''}$~$\geq$~9 were not detected.  The Doppler $b$-value ($b_{H2}$~=~(2$k$$T_{rot}$(H$_{2}$)/m$_{H2}$ + $v_{turb}^{2}$)$^{1/2}$,  in units of km s$^{-1}$), the radial velocity of the absorption lines (in units of km s$^{-1}$), and the molecular gas covering fraction are also free parameters in the fit. The normalized spectra (Figures 2 and 3) are smoothed with a 3 pixel boxcar average prior to H$_{2}$ fitting.  A synthetic normalized absorption spectrum is created for each combination of free parameters and this theoretical spectrum is convolved with the COS line-spread-function\footnote{The COS LSF experiences a wavelength dependent non-Gaussianity due to the introduction of mid-frequency wave-front errors produced by the polishing errors on the $HST$ primary and secondary mirrors; {\tt http://www.stsci.edu/hst/cos/documents/isrs/}}~\citep{kriss11}.  The process iterates until the best-fit parameters are found.  

Systematic uncertainties on the continuum normalization are explicitly taken into account during the fitting, and we execute an independent test to assess the magnitude of systematic uncertainties.  We vary parameters on the continuum normalization function until it deviates beyond the 1-sigma photometric error bars on the data.  The best-fit H$_{2}$ parameters are then re-derived and the difference between these and the nominal best-fit parameters are taken as the systematic errors. 

The spectral model comparison was carried out over the 1092.5~--~1117~\AA\ spectral window that covers most of the Lyman band absorptions from the (1~--~0) and (0~--~0) systems.  The individual best-fit column densities and statistical errors are shown in Table 1.  Spectral blending with $J^{''}$~=~3 states makes the column densities for $J^{''}$~=~2 and $J^{''}$~=~7 unreliable; the $J^{''}$~=~3 lines are reasonably well-constrained from the (1~--~0) band alone therefore these states are more robustly determined and dominate the fits in the 1106~--~1117~\AA\ region.  Consequently, we do not incorporate the $J^{''}$ = 2 and 7 lines in the subsequent analysis\footnote{including the $J^{''}$~=~2 and 7 column densities in the rotational temperature fit reduces the rotational temperature by $\lesssim$~10~\%.}.  The best-fit H$_{2}$ model spectrum is displayed in blue in Figures 2 and 3.  We find a total column of  log$_{10}$ $N(H_{2})$~=~19.90$^{+0.33}_{-0.22}$ cm$^{-2}$.  The uncertainty on the column density determination is a combination of statistical errors associated with the fitting process and systematic errors based on our choice of continuum normalization.  These errors are distributed as $\pm$~0.05 statistical fitting uncertainties and +0.33/$-$0.21 systematic uncertainties on the normalization.  The model finds a best fit $b$-value of 4~$\pm$~1 km s$^{-1}$, a radial velocity $v_{H2abs}$~=~5~$\pm$~5 km s$^{-1}$ and a covering fraction consistent with unity, $f_{cov}^{H2}$~=~0.974~$\pm$~0.027.

\begin{figure}
\begin{center}
\epsfig{figure=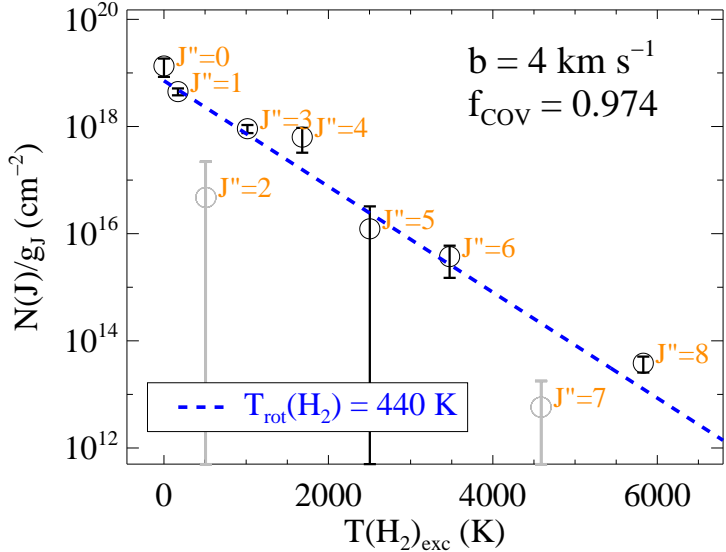,width=3.25in,angle=90}
\vspace{-0.25in}
\caption{
\label{cosovly} H$_{2}$ excitation diagram for the RW Aur disk absorption.  As described in the text, $J^{''}$ = 2 and 7 are poorly fit due to line blends and are not considered in the best fit excitation fit.  We find a best-fit rotational excitation temperature of the H$_{2}$  of $T_{rot}$~=~440~$\pm$~39 K.  
 }
\end{center}
\end{figure}

\begin{deluxetable}{cc}
\tabletypesize{\small}
\tablecaption{RW Aur A H$_{2}$ Parameters\tablenotemark{a}. \label{lya_lines}}
\tablewidth{0pt}
\tablehead{
\colhead{H$_{2}$ Level} & \colhead{log$_{10}$N(H$_{2}$, $v^{''}$~=~0, $J^{''}$) } 
}
\startdata
$J^{''}$~=~0	&        19.13~$\pm$~0.14 \\
$J^{''}$~=~1	&        19.61~$\pm$~0.06 \\
$J^{''}$~=~2	&        $\cdots$  \\
$J^{''}$~=~3	&        19.28~$\pm$~0.07 \\
$J^{''}$~=~4	&        18.76~$\pm$~0.17 \\
$J^{''}$~=~5	&        17.61~$\pm$~0.42 \\
$J^{''}$~=~6	&        16.69~$\pm$~0.20 \\
$J^{''}$~=~7	&        $\cdots$  \\
$J^{''}$~=~8	&        14.81~$\pm$~0.12 \\
\tableline
$T_{rot}$(H$_{2}$)  &  440~$\pm$~39 K \\ 
$b_{H2}$  &  4~$\pm$~1 km s$^{-1}$ \\ 
$f^{H2}_{cov}$  &   0.974~$\pm$~0.027 \\
\enddata
 \tablenotetext{a}{Column densities in each ground rotational state, $J^{''}$, are in units of molecules cm$^{-2}$.  Spectral fitting of $J^{''}$ = 2 and 7 levels are compromised by spectral blending at the resolution and S/N of the COS data. } 
\end{deluxetable}

 The critical densities for the low-$J^{''}$ rotational levels of H$_{2}$ are $\lesssim$~2~$\times$~10$^{4}$ cm$^{-3}$~\citep{mandy93}.  At the expected densities of the warm molecular disk layer ($n_{H2}$~$\sim$~10$^{6}$; Woitke et al. 2009; Bruderer 2013) we expect the H$_{2}$ level populations to be determined by collisions.  As such, the population will be described by the Maxwell-Boltzmann distribution with a form
\begin{equation}
N(J^{''})/N(J^{''}=0) = \frac{g_{J^{''}}}{g_{0}} e^{(-E_{J^{''}}/kT_{rot})} 
\end{equation}
where $T_{rot}$ is the rotational excitation temperature of the molecules, equal to the kinetic temperature ($T_{rot}$~=~$T_{kin}$) for a thermalized  population. Figure 5 displays  the H$_{2}$ excitation diagram for RW Aur A, from which a best fit $T_{rot}$(H$_{2}$)~=~440~$\pm$~39 K is derived.  Interestingly, a 440 K rotational temperature only corresponds to an H$_{2}$ thermal broadening of $\approx$~2 km s$^{-1}$, suggesting that a turbulent velocity of $v_{turb}$~$\sim$~3~--~3.5 km s$^{-1}$ could be present. However, this result is somewhat speculative because the strongest absorption lines are damped and therefore not highly sensitive to the exact $b$-value.  Higher S/N and spectral resolution measurements of H$_{2}$ absorption lines  should be able to provide more direct constraints on the turbulent broadening in inner disks. \nocite{woitke09,bruderer13} 

\subsection{CO Fitting}

The CO absorption line fitting and error estimation procedure is described in detail by~\citet{mcjunkin13}.  We summarize it briefly for the reader here: We analyze three bands of the CO Fourth Positive ($A$$^{1}\Pi$~--~$X$$^{1}\Sigma^{+}$)   system observed in our COS G160M spectra, (4~--~0), (2~--~0), and (1~--~0) with bandhead wavelengths of approximately 1419.0~\AA, 1477.6~\AA, and 1509.8~\AA, respectively.  Oscillator strengths and ground-state energy levels from the literature~\citep{haridass94,eidelsberg99,eidelsberg03} are used to compute synthetic CO absorption spectra.  We create a grid of model absorption spectra with the column densities of $^{12}$CO and $^{13}$CO ($N$($^{12}$CO) and $N$($^{13}$CO), in units of cm$^{-2}$), the CO rotational temperature $T_{rot}$(CO) (in units of Kelvin), the Doppler $b$-value ($b_{CO}$; in units of km s$^{-1}$), and the CO radial velocity ($v_{COabs}$; in units of km s$^{-1}$) as free parameters.  This grid is compared against the normalized CO absorption spectra to determine the best-fit CO parameters.  The
ranges of our grid search were 0.1~--~ 2.0 km s$^{-1}$ in steps of 0.1 km s$^{-1}$ for the Doppler $b$-value, 100~-–~1000 K with steps of 50 K for the rotational temperature, 14.0~-–~18.0 in steps of 0.1 for log$_{10}$(N($^{12}$CO)), and 14.0~--~17.0 cm$^{-2}$ in steps of 0.1 for log$_{10}$(N($^{13}$CO)). The maximum value of $b_{CO}$ is limited by our assumptions that $v_{turb}$~$<$ 1 km s$^{-1}$ and CO rotational temperatures $T_{rot}$(CO) $<$ 5~$\times$~10$^{3}$ K.  


The uncertainties on the model parameters were calculated from the photometric errors on the depth of the $^{12}$CO bandhead in the observed COS spectra.  The model parameter range was defined by varying each model parameter while keeping the other parameters constant; the best-fit parameter range was defined by models whose bandhead depths did not exceed the 1-sigma photometric error on the (2~--~0) bandhead depth in the data.

The best-fit CO parameters are log$_{10}$ N(CO)~=~16.1$^{+0.3}_{-0.5}$ cm$^{-2}$, $T_{rot}$(CO)~=~200$^{+650}_{-125}$ K, $b_{CO}$~=~0.5~$\pm$~0.1 km s$^{-1}$, and $v_{COabs}$~=~5~$\pm$~5 km s$^{-1}$. These results are consistent with the CO absorption parameters determined for RW Aur previously~\citep{mcjunkin13}.   The normalized CO data with the best fit model are shown in Figure 4.  We observe a possible blueshifted CO absorption component that may be associated with the low-ionization outflow, discussed in \S5.1.   The best fit $^{13}$CO column density is  log$_{10}$ $N$$^{13}$(CO)~=~14.3, however spectral blending with H$_{2}$ fluorescence lines and insufficient S/N prevent a meaningful determination of the $^{12}$CO/$^{13}$CO ratio from these data.


\begin{figure}
\begin{center}
\epsfig{figure=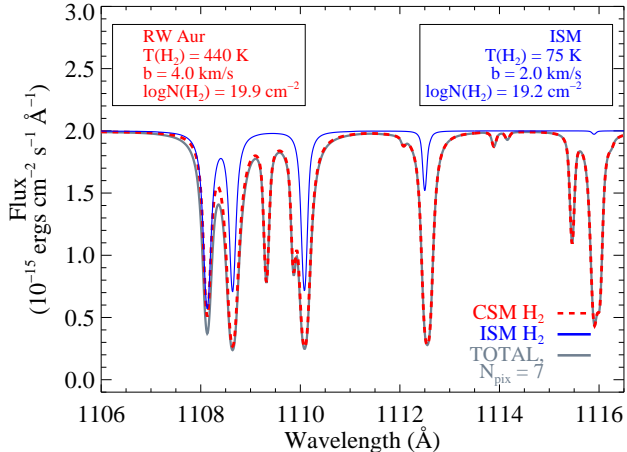,width=2.65in,angle=90}
\vspace{-0.25in}
\caption{
\label{cosovly} A comparison of model spectra for circumstellar H$_{2}$ (red dashed line) and interstellar H$_{2}$ (blue solid line) absorption systems shows that the circumstellar component dominates the total H$_{2}$ opacity.  The interstellar absorption spectrum is calculated from the interstellar atomic hydrogen and dust reddening values for the RW Aur line-of-sight (McJunkin et al. 2014; \S4.3).  
 }
\end{center}
\end{figure}

\subsection{Circumstellar Origin of the H$_{2}$ Absorption}

There is a rich literature on interstellar H$_{2}$ and CO absorption line studies of the sightlines to hot stars, from early sounding rocket observations~\citep{carruthers70}, to $Copernicus$~\citep{savage77,federman80}, though combined studies using $HST$ and $FUSE$~\citep{burgh07,sheffer08}.  Therefore, it is important to demonstrate that the molecular absorption observed towards RW Aur is indeed circumstellar and not interstellar.  This can be done with by comparing the properties of the H$_{2}$ and CO absorbers in RW Aur with those typical of the ISM.

The primary argument for the circumstellar origin of the observed H$_{2}$ is the observed rotational temperatures of the H$_{2}$ and CO populations.   The average H$_{2}$ rotational (kinetic) temperature in the diffuse and translucent ISM is $\approx$~60~--~100 K~\citep{savage77,rachford02} and the CO temperatures (typically sub-thermal in the ISM) are $<$ 10 K~\citep{burgh07,sheffer08}.  Therefore, the typical ISM temperatures of H$_{2}$ and CO are factors of $\sim$~6 and $\sim$~20 lower than observed for RW Aur, respectively.  The molecular rotational temperatures derived from our COS spectra are instead consistent with those expected for a warm molecular layer of a protoplanetary disk atmosphere~\citep{glassgold04,woitke09,meijerink12}.  

The measured molecular column densities are also much larger than would be expected based on the interstellar reddening towards RW Aur.  \citet{mcjunkin14} have recently presented direct measurements of the interstellar \ion{H}{1} column densities towards a sample of 31 young stars, including RW Aur.  For RW Aur, they measured log$_{10}$ N(HI) = 20.25$^{+0.05}_{-0.21}$ cm$^{-2}$ (the X-ray derived ``\ion{H}{1} column'' is a factor of 10 higher, however, the X-ray absorption is not a direct measurement of the neutral hydrogen column, see McJunkin et al. 2014).   Combining this with the well-characterized relationship between N(HI) and the selective reddening $E(B - V)$~\citep{diplas94}\footnote{If we instead adopt the calibration of the N(HI) vs. $E(B - V)$ relationship suggested by~\citet{liszt14}, the derived reddening is $\sim$~40~\% lower.}, we find $E(B - V)$~=~0.036.    The~\citet{bohlin78} relation can then be used to calculate the expected interstellar H$_{2}$ column density: 2N(H$_{2}$) = ($E(B - V)$~$\times$~(5.8~$\times$~10$^{21}$)~-~N(HI)).  This yields an interstellar H$_{2}$ column density of log$_{10}$ N$_{ISM}$(H$_{2}$)~=~19.19 cm$^{-2}$, a factor of approximately 5 lower than observed towards RW Aur.  This is shown graphically in Figure 6; we compare model absorption spectra of the ISM toward RW Aur (blue solid line) with the model fits to the observed data (red dashed line).  As one can see, only in the (0~--~0) R(0) $\lambda$ 1108.13~\AA\ line does the ISM  contribute appreciably to the opacity in     the observed spectrum.  The higher rotational states observed towards RW Aur are not predicted by the interstellar model.

An analogous argument can be made for the CO absorption lines.  \citet{burgh07} show the CO column density as a function of $E(B - V)$ for interstellar sightlines (their Figure 3, $left$).  For all sightlines with $E(B - V)$~$<$~0.2, they find log$_{10}$ N$_{ISM}$(CO)~$\leq$~13.3.    This is almost 10$^{3}$ times lower than the CO column density measured for RW Aur (\S4.2).  We conclude, based on both thermal and abundance arguments, that the H$_{2}$ and CO absorption line spectra observed towards RW Aur are completely dominated by circumstellar material.  

\begin{figure*}
\begin{center}
\epsfig{figure=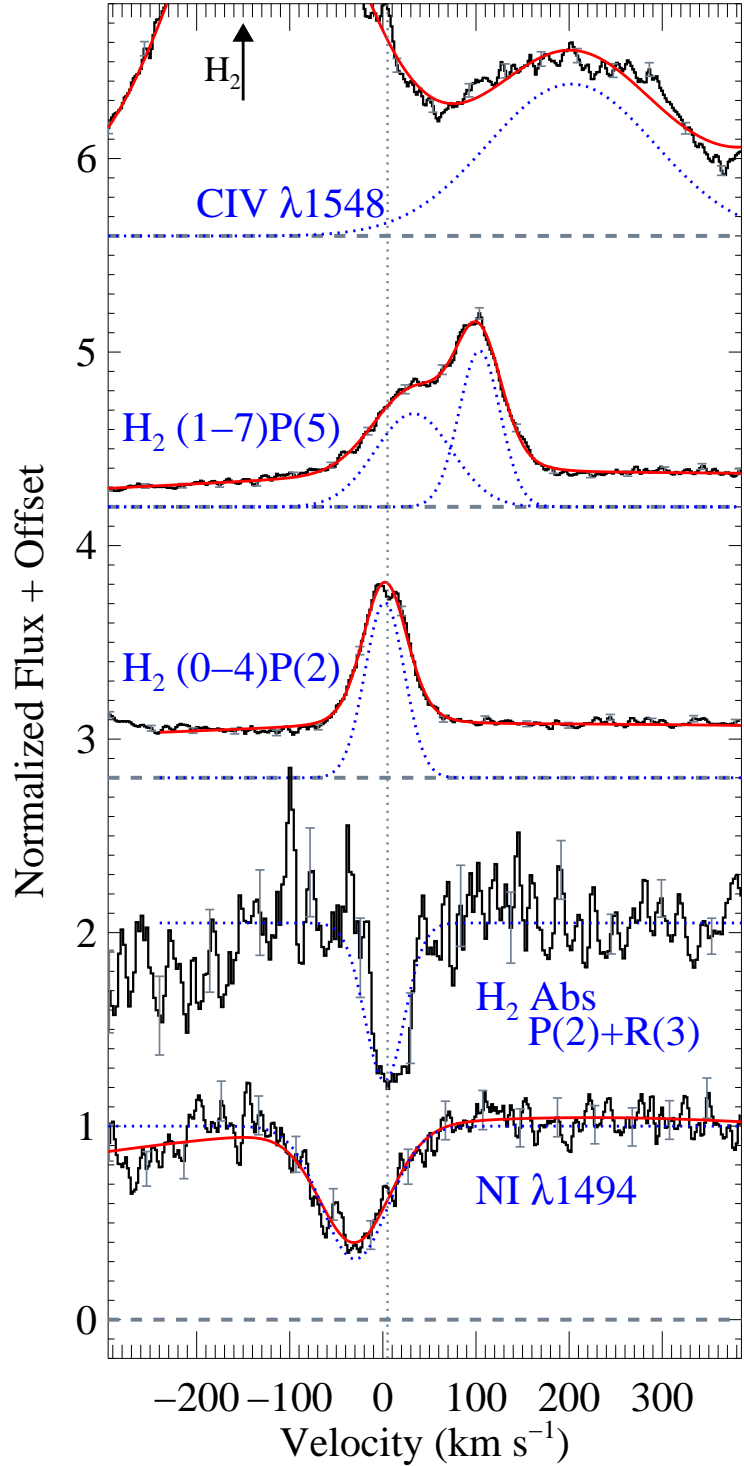,width=4.8in,angle=0}
\vspace{-0.1in}
\caption{
\label{cosovly} Emission and absorption profiles for species tracing all of the velocity fields in the 2013 RW Aur COS spectra (\S5.1, see also Table 2).  The line profiles are shown on a heliocentric velocity scale.  The profiles are normalized by the maximum (average continuum) flux in a given velocity interval for emission (absorption) lines, and Gaussian fits are shown as dotted blue lines to illustrate the velocity centroids.  From top, we show the red atomic lines (traced by \ion{C}{4}), the two-component red molecular outflow (traced by H$_{2}$ (1~--~7) P(5)), the molecular disk emission (traced by H$_{2}$ (0~--~4) P(2)) and absorption (traced by H$_{2}$ (0~--~0) P(2)), and the low-ionization blue-wind (traced by \ion{N}{1}).   
 }
\end{center}
\end{figure*}

\begin{figure}
\begin{center}
\epsfig{figure=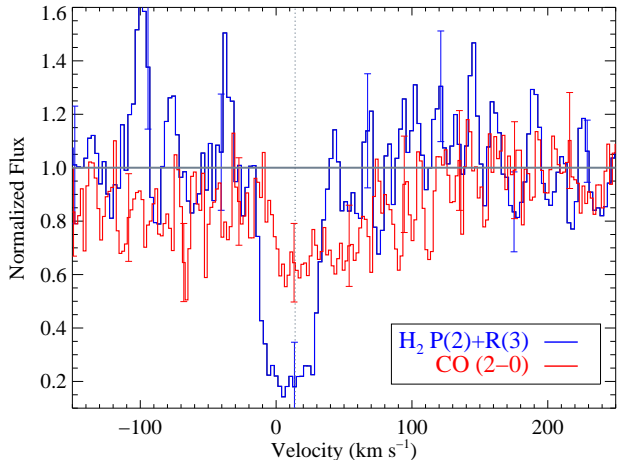,width=2.65in,angle=90}
\vspace{-0.25in}
\caption{
\label{cosovly} A closer look at the H$_{2}$ (blue) and CO (red) absorption velocities relative to the nominal +14 km s$^{-1}$ stellar radial velocity of RW Aur~\citep{hartmann86}.  At the $HST$-COS resolution, multiple H$_{2}$ and CO rotational lines are blended, but the average absorption velocity is consistent with the 4.7~$\mu$m CO emission velocity $v_{IR-CO}$~=~+9.5 km s$^{-1}$~\citep{brown13}.  
 }
\end{center}
\end{figure}

\begin{deluxetable*}{lcc}
\tabletypesize{\small}
\tablecaption{RW Aur A Velocity Components and Line Widths\tablenotemark{a}. \label{lya_lines}}
\tablewidth{0pt}
\tablehead{
\colhead{} & \colhead{$v_{rad}$ (km s$^{-1}$) } & 
\colhead{$FWHM$ (km s$^{-1}$) } 
}
\startdata
\tableline
Molecular Disk, $v$~$\approx$~0~--~+15 km s$^{-1}$ &   &   \\ 
\tableline
H$_{2}$ (0~--~0) absorption lines	& 0~--~+10 & $\cdots$  \\
CO (4,2,1~--~0) absorption lines		& 0~--~+10 & $\cdots$    \\
H$_{2}$ (0~--~4)P(2) emission		& +2.0~$\pm$~0.3 & 50~$\pm$~1   \\
H$_{2}$ (0~--~5)P(2) emission		& 13.9~$\pm$~1.0 & 45~$\pm$~2   \\
H$_{2}$ (4~--~1)R(17) emission		& $-$0.8~$\pm$~1.7 & 65~$\pm$~5   \\
H$_{2}$ (4~--~1)P(19) emission		& +1.9~$\pm$~0.7 & 62~$\pm$~2   \\
H$_{2}$ (4~--~4)R(17) emission		& +2.6~$\pm$~2.5 & 37~$\pm$~6   \\
H$_{2}$ (2~--~3)R(11) emission		& +11.5~$\pm$~0.7 & 48~$\pm$~2   \\
H$_{2}$ (2~--~6)R(11) emission		& +15.7~$\pm$~0.8 & 61~$\pm$~2   \\
H$_{2}$ (2~--~8)R(11) emission		& +7.9~$\pm$~0.8 & 49~$\pm$~2   \\
\tableline
Blue Wind\tablenotemark{b}, $v$~$\leq$~$-$30 km s$^{-1}$  &  &   \\ 
\tableline
\ion{Si}{3} $\lambda$1206 absorption  	&  $-$30.3 $\pm$ 0.8  &  292~$\pm$~8	   \\
\ion{N}{1} $\lambda$1243 absorption  	&  $-$79.9 $\pm$ 3.3  &  152~$\pm$~12	   \\
\ion{N}{1} $\lambda$1492 absorption  	&  $-$42.6 $\pm$ 1.2  &  100~$\pm$~4	   \\
\ion{N}{1} $\lambda$1494 absorption  	&  $-$29.8 $\pm$ 1.3  &  83~$\pm$~4	   \\
\ion{Si}{2} $\lambda$1526 absorption  	&  $-$28.9 $\pm$ 1.3  &  134~$\pm$~5	   \\
\ion{Si}{2} $\lambda$1533 absorption  	&  $-$37.5 $\pm$ 1.1  &  155~$\pm$~3	   \\
\tableline
Red Molecular Outflow, $v$~$\approx$~+40, +100 km s$^{-1}$ &  &   \\ 
\tableline
H$_{2}$ (1~--~6)P(5) emission$^{1}$		& +52.8~$\pm$~1.1  & 138~$\pm$~2   \\
H$_{2}$ (1~--~6)P(5) emission$^{2}$		& +109.6~$\pm$~0.5  & 47~$\pm$~2   \\

H$_{2}$ (1~--~7)R(3) emission$^{1}$		& +48.1~$\pm$~1.1  & 162~$\pm$~3   \\
H$_{2}$ (1~--~7)R(3) emission$^{2}$		& +104.1~$\pm$~0.5  & 44~$\pm$~2   \\

H$_{2}$ (1~--~7)P(5) emission$^{1}$		& +32.9~$\pm$~2.0  & 96~$\pm$~3   \\
H$_{2}$ (1~--~7)P(5) emission$^{2}$		& +103.8~$\pm$~0.6  & 52~$\pm$~1   \\

H$_{2}$ (1~--~3)R(6) emission$^{1}$		& +47.5~$\pm$~3.9  & 69~$\pm$~7   \\
H$_{2}$ (1~--~3)R(6) emission$^{2}$		& +103.9~$\pm$~5.0  & 57~$\pm$~6   \\

H$_{2}$ (1~--~7)R(6) emission$^{1}$		& +25.0~$\pm$~1.5  & 51~$\pm$~3   \\
H$_{2}$ (1~--~7)R(6) emission$^{2}$		& +97.0~$\pm$~1.0  & 61~$\pm$~6   \\

H$_{2}$ (1~--~7)P(8) emission$^{1}$		& +18.2~$\pm$~2.2  & 90~$\pm$~4   \\
H$_{2}$ (1~--~7)P(8) emission$^{2}$		& +102.0~$\pm$~0.8  & 61~$\pm$~3   \\

\tableline
Red Atomic Emission, $v$~$\geq$~+30 km s$^{-1}$ &  &   \\
\tableline
\ion{C}{4} $\lambda$1548 emission  	&  +203  &  211	   \\
\ion{C}{4} $\lambda$1550 emission  	&  +54  &  239	   \\
\ion{He}{2} $\lambda$1640 emission  	&  +90  & 145	   \\
\ion{O}{3}] $\lambda$1664 emission  	&  +39  & 341	   \\
\ion{N}{4}] $\lambda$1483 emission  	&  +92  & 212	   \\
\ion{N}{4}] $\lambda$1486 emission  	&  +2   & 367	   \\
\enddata
 \tablenotetext{a}{The $HST$-COS velocity scale has an accuracy of 15 km s$^{-1}$, the quoted velocity error bars are fitting uncertainties.  The line centers and FWHMs are fit assuming a Gaussian line shape that has been convolved with the local COS line-spread-function; the centroid and line widths presented here are those of the intrinsic Gaussian emission/absorption component, prior to instrumental convolution.  } 
\end{deluxetable*}


\section{Discussion}

\subsection{Spatially Decomposing the RW Aur A Environment: Velocity Fields}

As described in Section 2, the inner region of the RW Aur system is a dynamically active place.  Material being funneled onto the central star, a rotating circumstellar disk, and atomic/molecular outflows are present within~$\sim$~10~AU of the central star.  In this subsection, we decompose the kinematic signatures of five individual velocity fields, and show that the H$_{2}$ and CO absorption described above most likely originate in the circumstellar disk orbiting RW Aur A. We show representative lines from all five velocity fields in Figure 7.  A list of the observed spectral features in the COS spectra of RW Aur is presented in Table 2\footnote{We do not include measurements of all of the H$_{2}$ fluorescent emission lines here.  The general behavior is captured by the 14 lines we have featured.  Note that unlike many CTTSs (e.g., France et al. 2011; Schindhelm et al. 2012), CO fluorescence is not observed in the RW Aur spectrum.  ~\nocite{france11b,schindhelm12a}}.  The absolute velocity accuracy of
the COS FUV modes is $\approx$~15 km s$^{-1}$. 

\subsubsection{Molecular Disk} The first velocity field comprises molecular emission and absorptions between $\approx$~$-$1~--~+15 km 
  s$^{-1}$, and we attribute this to the molecular disk.  The canonical heliocentric stellar velocity of RW Aur A is $\approx$~+14 km s$^{-1}$, and while the variable nature of the star makes this number uncertain (e.g., Hartmann et al. 1986), it agrees well with radial velocity measurements of RW Aur A (15.87~$\pm$~0.55 km s$^{-1}$, but with a 5.7 km s$^{-1}$ radial velocity centroid oscillation, Gahm et al. 1999; 15.8 km s$^{-1}$, Cabrit et al. 2006) and RW Aur B (15.00~$\pm$~0.03 km s$^{-1}$; Nguyen et al. 2012).\nocite{nguyen12,najita03,gahm99}   
 The molecular inner disk of RW Aur A (traced by CO fundamental emission near 4.7~$\mu$m) is blue-shifted by $\approx$~7 km s$^{-1}$ relative to the typical stellar velocity~\citep{brown13}.\nocite{hartmann86}  Due to line-blending and the moderate-to-low S/N, it is hard to determine a precise velocity for the individual molecular absorption features in the COS spectra, however they are consistent with a centroid velocity between 0~--~+10 km s$^{-1}$ (Figure 8).   We conclude that the molecular absorption species studied here are consistent with both the stellar radial velocity and the CO fundamental emission velocity.  
The absorbing molecular material is consistent with a disk origin, but we cannot rule out a contribution from a slow, molecular disk wind (e.g., Brown et al. 2013).  However, in the slow disk wind picture~\citep{pontoppidan11}, the wind is essentially disk material that has acquired a tangential velocity component, and is distinct from material that has been launched as an outflow from the accretion flow region.  We therefore refer to the absorbing CO and H$_{2}$ gas as having a disk origin.  

H$_{2}$ fluorescent progressions pumped by Ly$\alpha$ photons significantly redward of Ly$\alpha$ line-center ($v_{pump}$~$>$+360 km s$^{-1}$ from 1215.67~\AA) are at similar velocities, $v_{disk}$(H$_{2}$)~=~+6.8~$\pm$~6.3 km s$^{-1}$, with symmetric line profiles.  H$_{2}$ progressions pumped near Ly$\alpha$ line-center are only found in the outflows (see below).  This suggests a scenario where the center of the intrinsic, broad Ly$\alpha$ emission profile generated near the stellar photosphere 
is removed through resonant scattering by neutral hydrogen atoms before these photons illuminate the disk surface.  The average FWHM of the disk-origin H$_{2}$ emission lines is $\langle$FWHM$\rangle_{disk}$~=~52~$\pm$~10 km s$^{-1}$, suggesting an inner H$_{2}$ disk radius~\citep{france12b,salyk11} of $R_{in}$(H$_{2}$)~$\approx$~0.5 AU (assuming M$_{*}$ = 1.2 M$_{\odot}$ and $i$ = 77\arcdeg).   The dispersion in the disk velocities are compatible with the internal consistency of the COS wavelength solution (15 km s$^{-1}$; Holland et al. 2014\footnote{http://www.stsci.edu/hst/cos/documents/handbooks/current/cos\_cover.html}).

\subsubsection{Blue Wind and Red Molecular Outflow} RW Aur is known to drive atomic and molecular 
outflows~\citep{hirth94,beck08,melnikov09,france12b}.  We observe two outflow structures: a blue-shifted atomic absorption component and a multi-component red-shifted molecular system.  The blue-shifted absorption features are low-ionization metals, with an average outflow velocity of $-$42 km s$^{-1}$ ($\pm$ 20 km s$^{-1}$ standard deviation). The redshifted H$_{2}$ structures were identified by~\citet{france12b}, but interestingly, earlier $HST$-GHRS observations of these lines did not show large red-shifted velocities~\citep{ardila02}.  The red-shifted H$_{2}$ fluorescence lines are exclusively pumped by Ly$\alpha$ photons within 100 km s$^{-1}$ of Ly$\alpha$ line-center, suggesting that the pumping spectrum is a narrow Ly$\alpha$ profile produced by shocks within the outflow itself~\citep{walter03,saucedo03,schneider13}.  We note that the red-shifted H$_{2}$ outflow profiles observed in  August 2013 have a somewhat different morphology to those seen in the first COS observations acquired in March 2011.  While the earlier COS observations showed a sawtooth morphology with a steadily rising profile from roughly $-$10~--~+100 km s$^{-1}$, the 2013 observations show two distinct emission components: a broad ($\langle$FWHM$\rangle_{out,broad}$~$\sim$~100 km s$^{-1}$) component near $v_{out,broad}$(H$_{2}$)~$\sim$~40 km s$^{-1}$ and a narrow ($\langle$FWHM$\rangle_{out,narrow}$~$\sim$~50 km s$^{-1}$) component near $v_{out,narrow}$(H$_{2}$)~$\sim$~100 km s$^{-1}$ (Figure 7, H$_{2}$ (1~--~7)P(5)).  It is possible that the lower-velocity broad H$_{2}$ outflow component is associated with the +44 km s$^{-1}$ near-IR H$_{2}$ outflow detected in RW Aur~\citep{beck08}.  Future studies of the molecular outflows from RW Aur would benefit from long-slit spectral observations.  

\subsubsection{Red Atomic Emission} Finally, we observe red-shifted emission from relatively high ionization species at a variety of velocities.  The centroid velocity of \ion{C}{4} $\lambda$1548 is compromised by a line-blend with the red-shifted H$_{2}$ (1~--~8) R(3) line, and in general the high-ionization species displayed non-Gaussian, asymmetric line profiles.  The \ion{C}{4} $\lambda$1550 and \ion{He}{2} $\lambda$1640 lines are redshifted by approximately +200 km s$^{-1}$ relative to the 2011 data described in detail by~\citet{ardila13}.  The \ion{O}{3}] $\lambda$1666 centroid has redshifted by $\sim$~+100 km s$^{-1}$ relative to 2011 and the \ion{N}{4}] semi-forbidden lines are only detected in 2013.  There is a strong, broad unidentified feature near 1472~\AA\ that was not present in the 2011 spectra (possibly {\ion{S}{1}; Herczeg et al. 2005).\nocite{herczeg05}  The primary purpose of characterizing these velocities is to establish the origin of the molecular absorption, the evolution of the hot gas lines and far-UV continuum will be addressed in a future work.

\subsection{Components of the Inner Molecular Disk}

Given the numerous emitting and absorbing species attributed to the inner region of the RW Aur A disk, we will take a moment to review our best estimate for where each of these lines are being formed:  The 4.7 $\mu$m CO fundamental emission~\citep{najita03,brown13} originates in the inner $\sim$~1 AU around the star, likely with additional emission from a molecular disk wind~\citep{pontoppidan11}.  This explains the very broad line widths without pronounced double-peaked line profiles that would be expected from a purely Keplerian rotating disk.  This infrared emitting CO gas is at a temperature of~$\sim$~1000~--~2000 K~\citep{najita03}, and there is an insufficient amount of high-temperature, high-density CO beyond the emitting region to produce significant self-absorption of the CO fundamental line profiles.   The narrow far-UV H$_{2}$ fluorescence emission originates in the $\sim$~2500 K molecular disk surface between $\sim$~0.5~--~5 AU~\citep{france12b,schindhelm12b}.  While the emitting region may extend several scale heights above the disk surface, the lowest rovibrational levels of these fluorescent cascades do experience self-absorption (e.g., Herczeg et al. 2004).\nocite{herczeg04} However, at the resolution of the COS observations, this is not observable in the line profiles.  The H$_{2}$ emission line widths are likely dominated by rotation, but outflows may contribute to the line cores and blue wings of the observed profiles.  In a strong outflow source like RW Aur A, H$_{2}$ fluorescent emission is also produced directly in the outflowing material as described in \S~5.1.2.

The H$_{2}$ and CO $absorptions$ that are the subject of this work originate near the surface of the inner molecular disk, but likely at lower scale heights and larger semi-major axes than the H$_{2}$ fluorescence emission from the disk. This gas is seen in absorption against the continuum emission from the accreting protostar.   This absorbing molecular gas is at a temperature of a few hundred degrees K and is likely concentrated in the warm molecular surface layer from ~$\sim$~2~--~10~AU.  This absorbing material may also produce CO fluorescence of Ly$\alpha$ photons in some CTTSs~\citep{france11b, schindhelm12a}, but this ultraviolet CO emission is not observed towards RW Aur.   We note again that this interpretation hinges upon a reasonably highly inclined disk.  The measured inclination of the inner disk of RW Aur A is $i$~=~77\arcdeg$^{+13}_{-15}$~\citep{eisner07}, however there is considerable dispersion on inclination estimates in the literature.  Refined estimates of the disk inclination would be helpful to solidify the origin of the UV molecular absorption.

\begin{deluxetable}{ccc}
\tabletypesize{\small}
\tablecaption{RW Aur A Disk Abundance Ratios\tablenotemark{a}. \label{lya_lines}}
\tablewidth{0pt}
\tablehead{
\colhead{Species} & \colhead{log$_{10}$N(X)}  &   \colhead{N(X)/N(H$_{total})$\tablenotemark{a}}
}
\startdata
\ion{H}{1}    &         $<$ 20.25$^{+0.05}_{-0.21}$  & $<$ 0.53  \\
H$_{2}$      &        19.90$^{+0.33}_{-0.22}$  &  $>$ 0.24  \\
CO   	     &        16.1$^{+0.3}_{-0.5}$                 &     $>$ 3.7~$\times$~10$^{-5}$ \\

\enddata
 \tablenotetext{a}{The majority of the neutral atomic hydrogen on the RW Aur A sightline is thought to be interstellar~\citep{mcjunkin14}, therefore relative abundance ratios are lower limits.  } 
\tablenotetext{b}{N(H$_{total}$)~=~N(\ion{H}{1}) + 2N(H$_{2}$)  }
\end{deluxetable}

\subsection{The CO/H$_{2}$ Ratio and Molecular Fraction in the Disk}

Combining the kinematic, thermal, and abundance analyses presented above, we have shown that the H$_{2}$ and CO absorptions are dominated  by molecular gas in the circumstellar disk around RW Aur A.  The molecular abundance ratio of this material is CO/H$_{2}$~=~1.6$^{+4.7}_{-1.3}$~$\times$~10$^{-4}$.  This value is consistent with the canonical ratio of 10$^{-4}$ assumed for the disk initial conditions and suggests 
that little chemical processing has occurred in the warm molecular surface layer of the inner disk. 
The inner disk CO/H$_{2}$ in RW Aur is significantly higher than the recently reported abundance ratio in the TW Hya disk, (0.1~--~3)~$\times$~10$^{-5}$~\citep{favre13}.  RW Aur is significantly younger than TW Hya, which may suggest a timescale for chemical evolution between 1~--~10 Myr.  Favre et al. (2013) argue that the low CO abundance in the TW Hya disk is the result of CO being sequestered into hydrocarbons or CO$_{2}$ by a slow X-ray driven He$^{+}$ chemistry.  Early work on the physical processes that drive this abundance evolution in the disk predict a characteristic timescale for gas phase CO reduction of~$\sim$~3~$\times$~10$^{6}$ yr~\citep{aikawa97}, meaning that RW Aur may not have undergone the same level of chemical processing as TW Hya.  It is worth noting that the TW Hya study employed proxies for both the H$_{2}$ and CO column densities (HD (1~--~0) and C$^{18}$O (2~--~1), respectively, although this is unlikely to impact the results significantly).   Additionally, the far-IR/mm measurements are sampling gas at a few tens of degrees K in the outer disk surface, whereas the $HST$ data presented here are sampling gas at several hundred degrees K that likely originates closer to the star.  
Finally, TW Hya is bright enough for the type of direct abundance analysis presented here, however the face-on geometry~\citep{qi06}  is not favorable for UV disk absorption observations (see Herczeg et al. 2004).\nocite{herczeg04} 

The total line-of-sight \ion{H}{1} column density towards RW Aur is log$_{10}$ N(HI) = 20.25$^{+0.05}_{-0.21}$ cm$^{-2}$, thought to be dominated by neutral hydrogen in the ISM.  This value sets an upper limit to the amount of neutral hydrogen on the circumstellar line-of-sight through the disk, $N_{CSM}$(HI)~$\leq$~$N_{ISM}$(HI); a lower limit to the molecular fraction of the warm disk atmosphere probed by our $HST$ measurements is $f_{H2}$~$\geq$~2$N$(H$_{2}$) / ($N_{ISM}$(HI) + 2$N$(H$_{2}$)).  We constrain the molecular fraction in the warm disk atmosphere of RW Aur to be $f_{H2}$~$\geq$~0.47.  This large molecular fraction is interesting because it has been suggested that circumstellar material with large molecular fractions may explain the discrepancy between optical/IR-based reddening values and $N$(HI)-based reddening values without having to invoke grain populations or gas-to-dust ratios that differ significantly from the diffuse and translucent ISM~\citep{mcjunkin14}.  Table 3 summarizes the atomic and molecular results from this work.  The combination of dense cloud CO/H$_{2}$ ratio and high molecular fraction argue that the warm molecular surface layers of protoplanetary disks retain some of the physical characteristics of the dense clouds out of which they formed, at least to the $\sim$~1 Myr age of RW Aur.

Finally, as described in the Introduction, observations of H$_{2}$ and CO UV fluorescence lines suggested high CO/H$_{2}$ ratios (0.1~--~1; e.g., France et al. 2011b; 2012).  While we caution that the results presented in the present study are based on a sample size of one, they suggest that the CO/H$_{2}$ ratios derived from the emission line studies were likely misleading because of the spatial stratification of the emitting regions being studied.  With a 2.5\arcsec\ diameter spectroscopic aperture, $HST$-COS emission line observations sample the entire inner disk surface ($r$~$\lesssim$~200 AU), where disk surface temperatures change from a few thousands to a few tens of degrees.  The combination of the derived rotational excitation temperatures and analyses of the rotationally broadened line widths argue that the $T_{rot}$(H$_{2}$)~$\sim$~2500 K gas originates inside 3 AU, while the cooler fluorescent CO emission originates between 2~--~10 AU~\citep{schindhelm12a,france12b}.   These regions are spatially unresolved by COS, so while the local column densities derived for each component are robust, the columns refer to different populations of gas and therefore the inferred local CO/H$_{2}$ ratio from spatially unresolved UV fluorescence observations is not meaningful.  

The UV fluorescent picture of the inner disk appears to be the following: the hot H$_{2}$-emitting population has too little CO associated with it to be detectable and the warm CO-emitting population is too cold for Ly$\alpha$ fluorescence of H$_{2}$ to operate efficiently.  The new H$_{2}$ and CO absorption line spectra presented in this work overcome these geometric complications by sampling material on a single pencil-beam sightline to the central star.   The similarity of the CO and H$_{2}$ excitation temperatures and velocity structure argue for a common spatial origin, with a CO/H$_{2}$ abundance ratio of CO/H$_{2}$~$\approx$~1.6~$\times$~10$^{-4}$.  

\section{Summary}   

We have presented new contemporaneous measurements of CO and H$_{2}$ absorption through the ``warm molecular layer'' in the protoplanetary disk around the Classical T Tauri Star RW Aurigae A.  We have demonstrated the use of a newly commissioned observing mode of the {\it Hubble Space Telescope} to detect warm H$_{2}$ in this region for the first time.  Spectral analyses of these data reveal the following major findings: 

\begin{enumerate}
\item The spectra are composed of emission from the accretion region near the stellar photosphere, the molecular disk, and several outflow components.   The relative spatial distribution of these components can be inferred from their velocities.  

\item Absorption spectra from H$_{2}$ and CO are observed and are consistent with an origin in the upper layers of the RW Aur circumstellar disk.  A low-ionization, low-velocity atomic outflow is also detected in absorption.  

\item Spectral synthesis modeling indicates that the molecular absorbers arise in a common parcel of disk gas, characterized by log$_{10}$ N(H$_{2}$)~=~19.90$^{+0.33}_{-0.22}$ cm$^{-2}$ at T$_{rot}$(H$_{2}$) ~=~440~$\pm$~39 K, with a molecular fraction $f_{H2}$~$\geq$~0.47.  The CO component has log$_{10}$ N(CO)~=~16.1~$^{+0.3}_{-0.5}$ cm$^{-2}$  at  T$_{rot}$(CO) ~=~200$^{+650}_{-125}$ K.  

\item We derive an abundance ratio of CO/H$_{2}$~=~1.6$^{+4.7}_{-1.3}$~$\times$~10$^{-4}$ for the inner disk gas, consistent with canonical interstellar dense cloud value.  

\end{enumerate}

\acknowledgments
The data presented here were obtained through $HST$ Guest Observing program 12876.  Initial design and characterization of the COS G130M $\lambda$1222 mode was performed as part of $HST$ Guest Observing program 12505. We appreciate helpful discussions with Eric Burgh, Eric Schindhelm, and Christian Schneider during the course of this work.  This work was partially supported by NASA grant NNX08AC146 to the University of Colorado at Boulder and KF acknowledges support from a Nancy Grace Roman Fellowship during a portion of this work. 

\bibliography{ms_H2CO_emapj}

\begin{thebibliography}{100}
\expandafter\ifx\csname natexlab\endcsname\relax\def\natexlab#1{#1}\fi

\bibitem[{{Aikawa} {et~al.}(1997){Aikawa}, {Umebayashi}, {Nakano}, \&
  {Miyama}}]{aikawa97}
{Aikawa}, Y., {Umebayashi}, T., {Nakano}, T., \& {Miyama}, S.~M. 1997, \apjl,
  486, L51

\bibitem[{{Andrews} \& {Williams}(2005)}]{andrews05}
{Andrews}, S.~M. \& {Williams}, J.~P. 2005, \apj, 631, 1134

\bibitem[{{Ardila} {et~al.}(2002){Ardila}, {Basri}, {Walter}, {Valenti}, \&
  {Johns-Krull}}]{ardila02}
{Ardila}, D.~R., {Basri}, G., {Walter}, F.~M., {Valenti}, J.~A., \&
  {Johns-Krull}, C.~M. 2002, \apj, 566, 1100

\bibitem[{{Ardila} {et~al.}(2013){Ardila}, {Herczeg}, {Gregory}, {Ingleby},
  {France}, {Brown}, {Edwards}, {Johns-Krull}, {Linsky}, {Yang}, {Valenti},
  {Abgrall}, {Alexander}, {Bergin}, {Bethell}, {Brown}, {Calvet}, {Espaillat},
  {Hillenbrand}, {Hussain}, {Roueff}, {Schindhelm}, \& {Walter}}]{ardila13}
{Ardila}, D.~R., {Herczeg}, G.~J., {Gregory}, S.~G., et al., 
  2013, \apjs, 207, 1

\bibitem[{{Armitage}(2007)}]{armitage07}
{Armitage}, P.~J. 2007, \apj, 665, 1381

\bibitem[{{Armitage} {et~al.}(2002){Armitage}, {Livio}, {Lubow}, \&
  {Pringle}}]{armitage02}
{Armitage}, P.~J., {Livio}, M., {Lubow}, S.~H., \& {Pringle}, J.~E. 2002,
  \mnras, 334, 248

\bibitem[{{Bary} {et~al.}(2008){Bary}, {Weintraub}, {Shukla}, {Leisenring}, \&
  {Kastner}}]{bary08}
{Bary}, J.~S., {Weintraub}, D.~A., {Shukla}, S.~J., {Leisenring}, J.~M., \&
  {Kastner}, J.~H. 2008, \apj, 678, 1088

\bibitem[{{Beck} {et~al.}(2008){Beck}, {McGregor}, {Takami}, \& {Pyo}}]{beck08}
{Beck}, T.~L., {McGregor}, P.~J., {Takami}, M., \& {Pyo}, T.-S. 2008, \apj,
  676, 472

\bibitem[{{Bergin} {et~al.}(2013){Bergin}, {Cleeves}, {Gorti}, {Zhang},
  {Blake}, {Green}, {Andrews}, {Evans}, {Henning}, {{\"O}berg}, {Pontoppidan},
  {Qi}, {Salyk}, \& {van Dishoeck}}]{bergin13}
{Bergin}, E.~A., {Cleeves}, L.~I., {Gorti}, U., et al.,  2013, \nat, 493, 644

\bibitem[{{Bohlin} {et~al.}(1978){Bohlin}, {Savage}, \& {Drake}}]{bohlin78}
{Bohlin}, R.~C., {Savage}, B.~D., \& {Drake}, J.~F. 1978, \apj, 224, 132

\bibitem[{{Bond} {et~al.}(2010){Bond}, {O'Brien}, \& {Lauretta}}]{bond10}
{Bond}, J.~C., {O'Brien}, D.~P., \& {Lauretta}, D.~S. 2010, \apj, 715, 1050

\bibitem[{{Brown} {et~al.}(2013){Brown}, {Pontoppidan}, {van Dishoeck},
  {Herczeg}, {Blake}, \& {Smette}}]{brown13}
{Brown}, J.~M., {Pontoppidan}, K.~M., {van Dishoeck}, E.~F., et al.,  2013, \apj, 770, 94

\bibitem[{{Bruderer}(2013)}]{bruderer13}
{Bruderer}, S. 2013, \aap, 559, A46

\bibitem[{{Burgh} {et~al.}(2007){Burgh}, {France}, \& {McCandliss}}]{burgh07}
{Burgh}, E.~B., {France}, K., \& {McCandliss}, S.~R. 2007, \apj, 658, 446

\bibitem[{{Cabrit} {et~al.}(2006){Cabrit}, {Pety}, {Pesenti}, \&
  {Dougados}}]{cabrit06}
{Cabrit}, S., {Pety}, J., {Pesenti}, N., \& {Dougados}, C. 2006, \aap, 452, 897

\bibitem[{{Carmona} {et~al.}(2008){Carmona}, {van den Ancker}, {Henning},
  {Pavlyuchenkov}, {Dullemond}, {Goto}, {Thi}, {Bouwman}, \&
  {Waters}}]{carmona08}
{Carmona}, A., {van den Ancker}, M.~E., {Henning}, T., et al.,  2008, \aap, 477, 839

\bibitem[{{Carruthers}(1970)}]{carruthers70}
{Carruthers}, G.~R. 1970, \apjl, 161, L81

\bibitem[{{Danforth} {et~al.}(2010){Danforth}, {Keeney}, {Stocke}, {Shull}, \&
  {Yao}}]{danforth10}
{Danforth}, C.~W., {Keeney}, B.~A., {Stocke}, J.~T., {Shull}, J.~M., \& {Yao},
  Y. 2010, \apj, 720, 976

\bibitem[{{Diplas} \& {Savage}(1994)}]{diplas94}
{Diplas}, A. \& {Savage}, B.~D. 1994, \apjs, 93, 211

\bibitem[{{Duch{\^e}ne} {et~al.}(1999){Duch{\^e}ne}, {Monin}, {Bouvier}, \&
  {M{\'e}nard}}]{duchene99}
{Duch{\^e}ne}, G., {Monin}, J.-L., {Bouvier}, J., \& {M{\'e}nard}, F. 1999,
  \aap, 351, 954

\bibitem[{{Dullemond} {et~al.}(2007){Dullemond}, {Hollenbach}, {Kamp}, \&
  {D'Alessio}}]{dullemond07}
{Dullemond}, C.~P., {Hollenbach}, D., {Kamp}, I., \& {D'Alessio}, P. 2007,
  Protostars and Planets V, 555

\bibitem[{{Eidelsberg} {et~al.}(1999){Eidelsberg}, {Jolly}, {Lemaire},
  {Tchang-Brillet}, {Breton}, \& {Rostas}}]{eidelsberg99}
{Eidelsberg}, M., {Jolly}, A., {Lemaire}, J.~L., {Tchang-Brillet}, W.-{\"U}.,
  {Breton}, J., \& {Rostas}, F. 1999, \aap, 346, 705

\bibitem[{{Eidelsberg} \& {Rostas}(2003)}]{eidelsberg03}
{Eidelsberg}, M. \& {Rostas}, F. 2003, \apjs, 145, 89

\bibitem[{{Eisner} {et~al.}(2007){Eisner}, {Hillenbrand}, {White}, {Bloom},
  {Akeson}, \& {Blake}}]{eisner07}
{Eisner}, J.~A., {Hillenbrand}, L.~A., {White}, R.~J., et al.,  2007, \apj, 669, 1072

\bibitem[{{Elias}(1978)}]{elias78}
{Elias}, J.~H. 1978, \apj, 224, 857

\bibitem[{{Fang} {et~al.}(2013){Fang}, {Kim}, {van Boekel}, {Sicilia-Aguilar},
  {Henning}, \& {Flaherty}}]{fang13}
{Fang}, M., {Kim}, J.~S., {van Boekel}, R., et al.,  2013, \apjs, 207, 5

\bibitem[{{Favre} {et~al.}(2013){Favre}, {Cleeves}, {Bergin}, {Qi}, \&
  {Blake}}]{favre13}
{Favre}, C., {Cleeves}, L.~I., {Bergin}, E.~A., {Qi}, C., \& {Blake}, G.~A.
  2013, \apjl, 776, L38

\bibitem[{{Fedele} {et~al.}(2010){Fedele}, {van den Ancker}, {Henning},
  {Jayawardhana}, \& {Oliveira}}]{fedele10}
{Fedele}, D., {van den Ancker}, M.~E., {Henning}, T., {Jayawardhana}, R., \&
  {Oliveira}, J.~M. 2010, \aap, 510, A72

\bibitem[{{Federman} {et~al.}(1980){Federman}, {Glassgold}, {Jenkins}, \&
  {Shaya}}]{federman80}
{Federman}, S.~R., {Glassgold}, A.~E., {Jenkins}, E.~B., \& {Shaya}, E.~J.
  1980, \apj, 242, 545

\bibitem[{{France} {et~al.}(2012{\natexlab{a}}){France}, {Burgh}, {Herczeg},
  {Schindhelm}, {Yang}, {Abgrall}, {Roueff}, {Brown}, {Brown}, \&
  {Linsky}}]{france12a}
{France}, K., {Burgh}, E.~B., {Herczeg}, G.~J., et al.,    2012{\natexlab{a}}, \apj, 744, 22

\bibitem[{{France} {et~al.}(2014){France}, {Schindhelm}, {Bergin}, {Roueff}, \&
  {Abgrall}}]{france14}
{France}, K., {Schindhelm}, E., {Bergin}, E.~A., {Roueff}, E., \& {Abgrall}, H.
  2014, \apj, 784, 127

\bibitem[{{France} {et~al.}(2011{\natexlab{a}}){France}, {Schindhelm}, {Burgh},
  {Herczeg}, {Harper}, {Brown}, {Green}, {Linsky}, {Yang}, {Abgrall}, {Ardila},
  {Bergin}, {Bethell}, {Brown}, {Calvet}, {Espaillat}, {Gregory},
  {Hillenbrand}, {Hussain}, {Ingleby}, {Johns-Krull}, {Roueff}, {Valenti}, \&
  {Walter}}]{france11b}
{France}, K., {Schindhelm}, E., {Burgh}, E.~B., et al.,  2011{\natexlab{a}}, \apj, 734, 31

\bibitem[{{France} {et~al.}(2012{\natexlab{b}}){France}, {Schindhelm},
  {Herczeg}, {Brown}, {Abgrall}, {Alexander}, {Bergin}, {Brown}, {Linsky},
  {Roueff}, \& {Yang}}]{france12b}
{France}, K., {Schindhelm}, E., {Herczeg}, G.~J., et al.,  2012{\natexlab{b}}, \apj, 756, 171

\bibitem[{{France} {et~al.}(2011{\natexlab{b}}){France}, {Yang}, \&
  {Linsky}}]{france11a}
{France}, K., {Yang}, H., \& {Linsky}, J.~L. 2011{\natexlab{b}}, \apj, 729, 7

\bibitem[{{Gahm} {et~al.}(1999){Gahm}, {Petrov}, {Duemmler}, {Gameiro}, \&
  {Lago}}]{gahm99}
{Gahm}, G.~F., {Petrov}, P.~P., {Duemmler}, R., {Gameiro}, J.~F., \& {Lago},
  M.~T.~V.~T. 1999, \aap, 352, L95

\bibitem[{{Glassgold} {et~al.}(2004){Glassgold}, {Najita}, \&
  {Igea}}]{glassgold04}
{Glassgold}, A.~E., {Najita}, J., \& {Igea}, J. 2004, \apj, 615, 972

\bibitem[{{Green} {et~al.}(2012){Green}, {Froning}, {Osterman}, {Ebbets},
  {Heap}, {Leitherer}, {Linsky}, {Savage}, {Sembach}, {Shull}, {Siegmund},
  {Snow}, {Spencer}, {Stern}, {Stocke}, {Welsh}, {B{\'e}land}, {Burgh},
  {Danforth}, {France}, {Keeney}, {McPhate}, {Penton}, {Andrews},
  {Brownsberger}, {Morse}, \& {Wilkinson}}]{green12}
{Green}, J.~C., {Froning}, C.~S., {Osterman}, S., et al.,  2012, \apj, 744, 60

\bibitem[{{Haridass} \& {Huber}(1994)}]{haridass94}
{Haridass}, C. \& {Huber}, K.~P. 1994, \apj, 420, 433

\bibitem[{{Hartigan} {et~al.}(1995){Hartigan}, {Edwards}, \&
  {Ghandour}}]{hartigan95}
{Hartigan}, P., {Edwards}, S., \& {Ghandour}, L. 1995, \apj, 452, 736

\bibitem[{{Hartigan} \& {Hillenbrand}(2009)}]{hartigan09}
{Hartigan}, P. \& {Hillenbrand}, L. 2009, \apj, 705, 1388

\bibitem[{{Hartmann} {et~al.}(1986){Hartmann}, {Hewett}, {Stahler}, \&
  {Mathieu}}]{hartmann86}
{Hartmann}, L., {Hewett}, R., {Stahler}, S., \& {Mathieu}, R.~D. 1986, \apj,
  309, 275

\bibitem[{{Hayashi} {et~al.}(1985){Hayashi}, {Nakazawa}, \&
  {Nakagawa}}]{hayashi85}
{Hayashi}, C., {Nakazawa}, K., \& {Nakagawa}, Y. 1985, in Protostars and
  planets II (A86-12626 03-90). Tucson, AZ, University of Arizona Press, 1985,
  p. 1100-1153., ed. {D.~C.~Black \& M.~S.~Matthews}, 1100--1153

\bibitem[{{Herczeg} \& {Hillenbrand}(2014)}]{herczeg14}
{Herczeg}, G.~J. \& {Hillenbrand}, L.~A. 2014, \apj, 786, 97

\bibitem[{{Herczeg} {et~al.}(2005){Herczeg}, {Walter}, {Linsky}, {Gahm},
  {Ardila}, {Brown}, {Johns-Krull}, {Simon}, \& {Valenti}}]{herczeg05}
{Herczeg}, G.~J., {Walter}, F.~M., {Linsky}, J.~L., et al.,   2005, \aj, 129, 2777

\bibitem[{{Herczeg} {et~al.}(2004){Herczeg}, {Wood}, {Linsky}, {Valenti}, \&
  {Johns-Krull}}]{herczeg04}
{Herczeg}, G.~J., {Wood}, B.~E., {Linsky}, J.~L., {Valenti}, J.~A., \&
  {Johns-Krull}, C.~M. 2004, \apj, 607, 369

\bibitem[{{Hern{\'a}ndez} {et~al.}(2007){Hern{\'a}ndez}, {Hartmann}, {Megeath},
  {Gutermuth}, {Muzerolle}, {Calvet}, {Vivas}, {Brice{\~n}o}, {Allen},
  {Stauffer}, {Young}, \& {Fazio}}]{hernandez07}
{Hern{\'a}ndez}, J., {Hartmann}, L., {Megeath}, T., et al.,  2007, \apj, 662, 1067

\bibitem[{{Hirth} {et~al.}(1994){Hirth}, {Mundt}, {Solf}, \& {Ray}}]{hirth94}
{Hirth}, G.~A., {Mundt}, R., {Solf}, J., \& {Ray}, T.~P. 1994, \apjl, 427, L99

\bibitem[{{Ida} \& {Lin}(2004)}]{ida04}
{Ida}, S. \& {Lin}, D.~N.~C. 2004, \apj, 604, 388

\bibitem[{{Ingleby} {et~al.}(2013){Ingleby}, {Calvet}, {Herczeg}, {Blaty},
  {Walter}, {Ardila}, {Alexander}, {Edwards}, {Espaillat}, {Gregory},
  {Hillenbrand}, \& {Brown}}]{ingleby13}
{Ingleby}, L., {Calvet}, N., {Herczeg}, G., et al.,  2013, \apj, 767, 112

\bibitem[{{Ingleby} {et~al.}(2011{\natexlab{a}}){Ingleby}, {Calvet},
  {Hern{\'a}ndez}, {Brice{\~n}o}, {Espaillat}, {Miller}, {Bergin}, \&
  {Hartmann}}]{ingleby11b}
{Ingleby}, L., {Calvet}, N., {Hern{\'a}ndez}, J., et al., 
  2011{\natexlab{a}}, \aj, 141, 127

\bibitem[{{Ingleby} {et~al.}(2011{\natexlab{b}}){Ingleby}, {Calvet},
  {Hern{\'a}ndez}, {Brice{\~n}o}, {Espaillat}, {Miller}, {Bergin}, \&
  {Hartmann}}]{ingleby11a}
---. 2011{\natexlab{b}}, \aj, 141, 127

\bibitem[{{Jayawardhana} {et~al.}(2006){Jayawardhana}, {Coffey}, {Scholz},
  {Brandeker}, \& {van Kerkwijk}}]{rayjay06}
{Jayawardhana}, R., {Coffey}, J., {Scholz}, A., {Brandeker}, A., \& {van
  Kerkwijk}, M.~H. 2006, \apj, 648, 1206

\bibitem[{{Kenyon} \& {Hartmann}(1995)}]{kenyon95}
{Kenyon}, S.~J. \& {Hartmann}, L. 1995, \apjs, 101, 117

\bibitem[{{Kriss}(2011)}]{kriss11}
{Kriss}, G.~A. 2011, {Improved Medium Resolution Line Spread Functions for COS
  FUV Spectra}, Tech. rep.

\bibitem[{{Lacy} {et~al.}(1994){Lacy}, {Knacke}, {Geballe}, \&
  {Tokunaga}}]{lacy94}
{Lacy}, J.~H., {Knacke}, R., {Geballe}, T.~R., \& {Tokunaga}, A.~T. 1994,
  \apjl, 428, L69

\bibitem[{{Lahuis} {et~al.}(2007){Lahuis}, {van Dishoeck}, {Blake}, {Evans},
  {Kessler-Silacci}, \& {Pontoppidan}}]{lahuis07}
{Lahuis}, F., {van Dishoeck}, E.~F., {Blake}, G.~A., et al.,  2007, \apj, 665, 492

\bibitem[{{Liszt}(2014)}]{liszt14}
{Liszt}, H. 2014, \apj, 780, 10

\bibitem[{{Mandy} \& {Martin}(1993)}]{mandy93}
{Mandy}, M.~E. \& {Martin}, P.~G. 1993, \apjs, 86, 199

\bibitem[{{Markwardt}(2009)}]{markwardt09}
{Markwardt}, C.~B. 2009, in Astronomical Society of the Pacific Conference
  Series, Vol. 411, Astronomical Data Analysis Software and Systems XVIII, ed.
  D.~A. {Bohlender}, D.~{Durand}, \& P.~{Dowler}, 251

\bibitem[{{Martin-Za{\"i}di} {et~al.}(2010){Martin-Za{\"i}di}, {Augereau},
  {M{\'e}nard}, {Olofsson}, {Carmona}, {Pinte}, \& {Habart}}]{zaidi10}
{Martin-Za{\"i}di}, C., {Augereau}, J.-C., et al.,  2010, \aap, 516, A110

\bibitem[{{Martin-Za{\"i}di} {et~al.}(2008){Martin-Za{\"i}di}, {Deleuil}, {Le
  Bourlot}, {Bouret}, {Roberge}, {Dullemond}, {Testi}, {Feldman}, {Lecavelier
  Des Etangs}, \& {Vidal-Madjar}}]{zaidi08}
{Martin-Za{\"i}di}, C., {Deleuil}, M., {Le Bourlot}, J., et al.,  2008, \aap, 484, 225

\bibitem[{{McCandliss}(2003)}]{mccandliss03}
{McCandliss}, S.~R. 2003, \pasp, 115, 651

\bibitem[{{McCandliss} {et~al.}(2010){McCandliss}, {France}, {Osterman},
  {Green}, {McPhate}, \& {Wilkinson}}]{mccandliss10}
{McCandliss}, S.~R., {France}, K., {Osterman}, S., et al.,  2010, \apjl, 709, L183

\bibitem[{{McJunkin} {et~al.}(2013){McJunkin}, {France}, {Burgh}, {Herczeg},
  {Schindhelm}, {Brown}, \& {Brown}}]{mcjunkin13}
{McJunkin}, M., {France}, K., {Burgh}, E.~B., et al.,  2013, \apj, 766, 12

\bibitem[{{McJunkin} {et~al.}(2014){McJunkin}, {France}, {Schneider},
  {Herczeg}, {Brown}, {Hillenbrand}, {Schindhelm}, \& {Edwards}}]{mcjunkin14}
{McJunkin}, M., {France}, K., {Schneider}, P.~C., et al.,  2014, \apj, 780,
  150

\bibitem[{{Meijerink} {et~al.}(2012){Meijerink}, {Aresu}, {Kamp}, {Spaans},
  {Thi}, \& {Woitke}}]{meijerink12}
{Meijerink}, R., {Aresu}, G., {Kamp}, I., et al.,  2012, \aap, 547, A68

\bibitem[{{Melnikov} {et~al.}(2009){Melnikov}, {Eisl{\"o}ffel}, {Bacciotti},
  {Woitas}, \& {Ray}}]{melnikov09}
{Melnikov}, S.~Y., {Eisl{\"o}ffel}, J., {Bacciotti}, F., {Woitas}, J., \&
  {Ray}, T.~P. 2009, \aap, 506, 763

\bibitem[{{Mordasini} {et~al.}(2009){Mordasini}, {Alibert}, \&
  {Benz}}]{mordasini09}
{Mordasini}, C., {Alibert}, Y., \& {Benz}, W. 2009, \aap, 501, 1139

\bibitem[{{Najita} {et~al.}(2003){Najita}, {Carr}, \& {Mathieu}}]{najita03}
{Najita}, J., {Carr}, J.~S., \& {Mathieu}, R.~D. 2003, \apj, 589, 931

\bibitem[{{Najita} {et~al.}(2007){Najita}, {Carr}, {Glassgold}, \&
  {Valenti}}]{najita07}
{Najita}, J.~R., {Carr}, J.~S., {Glassgold}, A.~E., \& {Valenti}, J.~A. 2007,
  Protostars and Planets V, 507

\bibitem[{{Nguyen} {et~al.}(2012){Nguyen}, {Brandeker}, {van Kerkwijk}, \&
  {Jayawardhana}}]{nguyen12}
{Nguyen}, D.~C., {Brandeker}, A., {van Kerkwijk}, M.~H., \& {Jayawardhana}, R.
  2012, \apj, 745, 119

\bibitem[{{{\"O}berg} {et~al.}(2011){{\"O}berg}, {Murray-Clay}, \&
  {Bergin}}]{oberg11}
{{\"O}berg}, K.~I., {Murray-Clay}, R., \& {Bergin}, E.~A. 2011, \apjl, 743, L16

\bibitem[{{Pascucci} {et~al.}(2006){Pascucci}, {Gorti}, {Hollenbach}, {Najita},
  {Meyer}, {Carpenter}, {Hillenbrand}, {Herczeg}, {Padgett}, {Mamajek},
  {Silverstone}, {Schlingman}, {Kim}, {Stobie}, {Bouwman}, {Wolf}, {Rodmann},
  {Hines}, {Lunine}, \& {Malhotra}}]{pascucci06}
{Pascucci}, I., {Gorti}, U., {Hollenbach}, D., et al.,  2006, \apj, 651, 1177

\bibitem[{{Penton} {et~al.}(2013){Penton}, {Aloisi}, {Bostroem}, {Elliott},
  {France}, {Lockwood}, {Oliveira}, {Osterman}, {Proffitt}, {Roman-Duval}, \&
  {Sahnow}}]{penton13}
{Penton}, S.~V., {Aloisi}, A., {Bostroem}, K.~A., et al.,  2013, in American Astronomical Society
  Meeting Abstracts, Vol. 221, American Astronomical Society Meeting Abstracts,
  \#344.04

\bibitem[{{Penton} {et~al.}(2012){Penton}, {Osterman}, {France}, {Oliveira}, \&
  {Sahnow}}]{penton12}
{Penton}, S.~V., {Osterman}, S.~N., {France}, K., {Oliveira}, C., \& {Sahnow},
  D.~J. 2012, in American Astronomical Society Meeting Abstracts, Vol. 219,
  American Astronomical Society Meeting Abstracts \#219, \#241.19

\bibitem[{{Pontoppidan} {et~al.}(2011){Pontoppidan}, {Blake}, \&
  {Smette}}]{pontoppidan11}
{Pontoppidan}, K.~M., {Blake}, G.~A., \& {Smette}, A. 2011, \apj, 733, 84

\bibitem[{{Qi} {et~al.}(2006){Qi}, {Wilner}, {Calvet}, {Bourke}, {Blake},
  {Hogerheijde}, {Ho}, \& {Bergin}}]{qi06}
{Qi}, C., {Wilner}, D.~J., {Calvet}, N., et al.,  2006, \apjl, 636, L157

\bibitem[{{Rachford} {et~al.}(2002){Rachford}, {Snow}, {Tumlinson}, {Shull},
  {Blair}, {Ferlet}, {Friedman}, {Gry}, {Jenkins}, {Morton}, {Savage},
  {Sonnentrucker}, {Vidal-Madjar}, {Welty}, \& {York}}]{rachford02}
{Rachford}, B.~L., {Snow}, T.~P., {Tumlinson}, J., et al.,  2002, \apj, 577, 221

\bibitem[{{Richter} {et~al.}(2002){Richter}, {Jaffe}, {Blake}, \&
  {Lacy}}]{richter02}
{Richter}, M.~J., {Jaffe}, D.~T., {Blake}, G.~A., \& {Lacy}, J.~H. 2002, \apjl,
  572, L161

\bibitem[{{Roberge} {et~al.}(2001){Roberge}, {Lecavelier des Etangs}, {Grady},
  {Vidal-Madjar}, {Bouret}, {Feldman}, {Deleuil}, {Andre}, {Boggess},
  {Bruhweiler}, {Ferlet}, \& {Woodgate}}]{roberge01}
{Roberge}, A., {Lecavelier des Etangs}, A., {Grady}, C.~A.,et al.,  2001, \apjl, 551, L97

\bibitem[{{Salyk} {et~al.}(2009){Salyk}, {Blake}, {Boogert}, \&
  {Brown}}]{salyk09}
{Salyk}, C., {Blake}, G.~A., {Boogert}, A.~C.~A., \& {Brown}, J.~M. 2009, \apj,
  699, 330

\bibitem[{{Salyk} {et~al.}(2011){Salyk}, {Pontoppidan}, {Blake}, {Najita}, \&
  {Carr}}]{salyk11}
{Salyk}, C., {Pontoppidan}, K.~M., {Blake}, G.~A., {Najita}, J.~R., \& {Carr},
  J.~S. 2011, \apj, 731, 130

\bibitem[{{Saucedo} {et~al.}(2003){Saucedo}, {Calvet}, {Hartmann}, \&
  {Raymond}}]{saucedo03}
{Saucedo}, J., {Calvet}, N., {Hartmann}, L., \& {Raymond}, J. 2003, \apj, 591,
  275

\bibitem[{{Savage} {et~al.}(1977){Savage}, {Bohlin}, {Drake}, \&
  {Budich}}]{savage77}
{Savage}, B.~D., {Bohlin}, R.~C., {Drake}, J.~F., \& {Budich}, W. 1977, \apj,
  216, 291

\bibitem[{{Schindhelm} {et~al.}(2012{\natexlab{a}}){Schindhelm}, {France},
  {Burgh}, {Herczeg}, {Green}, {Brown}, {Brown}, \& {Valenti}}]{schindhelm12a}
{Schindhelm}, E., {France}, K., {Burgh}, et al.,  2012{\natexlab{a}},
  \apj, 746, 97

\bibitem[{{Schindhelm} {et~al.}(2012{\natexlab{b}}){Schindhelm}, {France},
  {Herczeg}, {Bergin}, {Yang}, {Brown}, {Brown}, {Linsky}, \&
  {Valenti}}]{schindhelm12b}
{Schindhelm}, E., {France}, K., {Herczeg}, G.~J., et al.,   2012{\natexlab{b}}, \apjl, 756, L23

\bibitem[{{Schneider} {et~al.}(2013){Schneider}, {Eisl{\"o}ffel}, {G{\"u}del},
  {G{\"u}nther}, {Herczeg}, {Robrade}, \& {Schmitt}}]{schneider13}
{Schneider}, P.~C., {Eisl{\"o}ffel}, J., {G{\"u}del}, M., et al.,  2013, \aap, 557, A110

\bibitem[{{Sheffer} {et~al.}(2008){Sheffer}, {Rogers}, {Federman}, {Abel},
  {Gredel}, {Lambert}, \& {Shaw}}]{sheffer08}
{Sheffer}, Y., {Rogers}, M., {Federman}, S.~R., et al.,  2008, \apj, 687, 1075

\bibitem[{{Sicilia-Aguilar} {et~al.}(2005){Sicilia-Aguilar}, {Hartmann},
  {Hern{\'a}ndez}, {Brice{\~n}o}, \& {Calvet}}]{aguilar05}
{Sicilia-Aguilar}, A., {Hartmann}, L.~W., {Hern{\'a}ndez}, J., {Brice{\~n}o},
  C., \& {Calvet}, N. 2005, \aj, 130, 188

\bibitem[{{Thi} {et~al.}(2004){Thi}, {van Zadelhoff}, \& {van
  Dishoeck}}]{thi04}
{Thi}, W.-F., {van Zadelhoff}, G.-J., \& {van Dishoeck}, E.~F. 2004, \aap, 425,
  955

\bibitem[{{Torres} {et~al.}(2007){Torres}, {Loinard}, {Mioduszewski}, \&
  {Rodr{\'{\i}}guez}}]{loinard07}
{Torres}, R.~M., {Loinard}, L., {Mioduszewski}, A.~J., \& {Rodr{\'{\i}}guez},
  L.~F. 2007, \apj, 671, 1813

\bibitem[{{Trilling} {et~al.}(2002){Trilling}, {Lunine}, \&
  {Benz}}]{trilling02}
{Trilling}, D.~E., {Lunine}, J.~I., \& {Benz}, W. 2002, \aap, 394, 241

\bibitem[{{Visser} {et~al.}(2011){Visser}, {Doty}, \& {van
  Dishoeck}}]{visser11}
{Visser}, R., {Doty}, S.~D., \& {van Dishoeck}, E.~F. 2011, \aap, 534, A132

\bibitem[{{Walter} {et~al.}(2003){Walter}, {Herczeg}, {Brown}, {Ardila},
  {Gahm}, {Johns-Krull}, {Lissauer}, {Simon}, \& {Valenti}}]{walter03}
{Walter}, F.~M., {Herczeg}, G., {Brown}, A.,et al.,   2003, \aj, 126, 3076

\bibitem[{{Ward}(1997)}]{ward97}
{Ward}, W.~R. 1997, Icarus, 126, 261

\bibitem[{{White} \& {Ghez}(2001)}]{white01}
{White}, R.~J. \& {Ghez}, A.~M. 2001, \apj, 556, 265

\bibitem[{{Woitas} {et~al.}(2001){Woitas}, {Leinert}, \&
  {K{\"o}hler}}]{woitas01}
{Woitas}, J., {Leinert}, C., \& {K{\"o}hler}, R. 2001, \aap, 376, 982

\bibitem[{{Woitas} {et~al.}(2002){Woitas}, {Ray}, {Bacciotti}, {Davis}, \&
  {Eisl{\"o}ffel}}]{woitas02}
{Woitas}, J., {Ray}, T.~P., {Bacciotti}, F., {Davis}, C.~J., \&
  {Eisl{\"o}ffel}, J. 2002, \apj, 580, 336

\bibitem[{{Woitke} {et~al.}(2009){Woitke}, {Dent}, {Thi}, {Sibthorpe}, {Rice},
  {Williams}, {Sicilia-Aguilar}, {Brown}, {Kamp}, {Pascucci}, {Alexander}, \&
  {Roberge}}]{woitke09}
{Woitke}, P., {Dent}, B., {Thi}, W., et al.,  2009, in American Institute of Physics
  Conference Series, Vol. 1094, American Institute of Physics Conference
  Series, ed. {E.~Stempels}, 225--233

\bibitem[{{Yang} {et~al.}(2011){Yang}, {Linsky}, \& {France}}]{yang11}
{Yang}, H., {Linsky}, J.~L., \& {France}, K. 2011, \apjl, 730, L10+

\end{thebibliography}



\end{document}